\documentclass[fleqn,usenatbib]{mnras}

\usepackage[T1]{fontenc}

\DeclareRobustCommand{\VAN}[3]{#2}
\let\VANthebibliography\thebibliography
\def\thebibliography{\DeclareRobustCommand{\VAN}[3]{##3}\VANthebibliography}

\usepackage{graphicx}	
\usepackage{amsmath}	
\usepackage{amssymb}	
\usepackage{newtxtext,newtxmath}

\usepackage{units}

\def\app#1#2{%
  \mathrel{%
    \setbox0=\hbox{$#1\sim$}%
    \setbox2=\hbox{%
      \rlap{\hbox{$#1\propto$}}%
      \lower1.1\ht0\box0%
    }%
    \raise0.25\ht2\box2%
  }%
}
\def\approxprop{\mathpalette\app\relax}

\title[Pulsar-Driven SN 1054]{SN 1054 as a Pulsar-Driven Supernova: Implications for the Crab Pulsar and Remnant Evolution}

\author[Omand et al.]{
Conor M. B. Omand \href{https://orcid.org/0000-0002-9646-8710}{\includegraphics[scale=0.5]{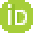}},$^{1,2}$\thanks{E-mail: c.m.omand@ljmu.ac.uk}
Nikhil Sarin \href{https://orcid.org/0000-0003-2700-1030}{\includegraphics[scale=0.5]{ORCIDiD_icon16x16.eps}},$^{3,4}$ 
and Tea Temim \href{https://orcid.org/0000-0001-7380-3144}{\includegraphics[scale=0.5]{ORCIDiD_icon16x16.eps}}$^{5}$ 
\\
$^{1}$Astrophysics Research Institute, Liverpool John Moores University, Liverpool Science Park IC2, 146 Brownlow Hill, Liverpool, UK, L3 5RF\\
$^{2}$The Oskar Klein Centre, Department of Astronomy, Stockholm University, AlbaNova, SE-106 91 Stockholm, Sweden\\
$^{3}$The Oskar Klein Centre, Department of Physics, Stockholm University, AlbaNova, SE-106 91 Stockholm, Sweden\\
$^{4}$Nordita,
Stockholm University and KTH Royal Institute of Technology
Hannes Alfvéns väg 12, SE-106 91 Stockholm, Sweden\\
$^{5}$ Department of Astrophysical Sciences, Princeton University, 4 Ivy Lane, Princeton, NJ 08544, USA
}

\date{Accepted XXX. Received YYY; in original form ZZZ}

\pubyear{2024}

\begin{document}
\label{firstpage}
\pagerange{\pageref{firstpage}--\pageref{lastpage}}
\maketitle

\begin{abstract}
One of the most studied objects in astronomy, the Crab Nebula, is the remnant of the historical supernova SN 1054.  Historical observations of the supernova imply a typical supernova luminosity, but contemporary observations of the remnant imply a low explosion energy and low ejecta kinetic energy.  These observations are incompatible with a standard $^{56}$Ni-powered supernova, hinting at an an alternate power source such as circumstellar interaction or a central engine.  We examine SN 1054 using a pulsar-driven supernova model, similar to those used for superluminous supernovae.  The model can reproduce the luminosity and velocity of SN 1054 for an initial spin period of $\sim$ 14 ms and an initial dipole magnetic field of 10$^{14-15}$ G.  We discuss the implications of these results, including the evolution of the Crab pulsar, the evolution of the remnant structure, formation of filaments, and limits on freely expanding ejecta. We discuss how our model could be tested further through potential light echo photometry and spectroscopy, as well as the modern analogues of SN 1054.
\end{abstract}

\begin{keywords}
supernovae: individual: SN 1054 -- pulsars: individual: Crab pulsar -- stars: magnetars-- ISM: supernova remnants
\end{keywords}



\section{Introduction}

The Crab Nebula is one of the most well-studied astronomical objects in the sky \citep[e.g.][and references therein]{Davidson1985, Hester2008, Buhler2014}.  It is one of a few remnants where the supernova (SN 1054) is recorded in historical records, and thus the age of the remnant is well constrained \citep{Clark1977}.  However, despite extensive studies, many questions still remain, such as the progenitor of the explosion and the explosion mechanism.  These questions are mostly driven by the unusually low kinetic energy inferred from studies of the remnant \citep{MacAlpine1989, Bietenholz1991, Fesen1997, Smith2003} despite the supernova being consistent with the luminosity of typical supernovae.

SN 1054 was observed for around two years by astronomers in Japan, China, and parts of Europe \citep{Clark1977, Collins1999}.  From records in China and Japan, the supernova was visible during the day for 23 days and during the night for around 650 days \citep{Clark1977}, although some European records suggest the supernova may have been bright enough to see during the day for several months \citep{Collins1999}.  The presence of a pulsar and detection of several solar masses of material in filaments makes it clear that SN 1054 was a core-collapse supernova, and the detection of a substantial amount of hydrogen in the filaments \citep{Davidson1985} implies a Type II classification~\citep{Dessart2012, Hachinger2012}.

The low distance to the Crab has allowed detailed observations of the structure of the Crab system, which consists of the pulsar, synchrotron nebula, thermal filaments, and freely expanding ejecta \citep{Hester2008}.  The observed velocities of the filaments range between $\sim$ 700 -- 1800 km s$^{-1}$, with a characteristic value of $\sim$ 1500 km s$^{-1}$ \citep{Clark1983, Bietenholz1991, Temim2006}.  These filaments show complex structures that likely arise from Rayleigh-Taylor instabilities at the interface between the synchrotron nebula and ejecta \citep{Davidson1985, Hester2008}.  The inferred kinetic energy of these filaments is $\lesssim$ 10$^{50}$ erg.  The freely expanding ejecta beyond the edge of the easily visible nebula was detected between $\sim$ 1200 -- 2500 km s$^{-1}$ in C IV $\lambda$1550 absorption \citep{Sollerman2000}, although no forward shock has been detected in either radio or X-ray beyond the edge of the synchrotron nebula and filaments \citep{Mauche1989, Predehl1995, Frail1995, Seward2006}.  The material detected by C IV absorption is consistent with a kinetic energy of 10$^{51}$ erg, but only for shallow density profiles \citep{Sollerman2000, Hester2008}.

What could have powered the unusually bright supernova luminosity?  The low inferred kinetic energy of the filaments has led to suggestions of SN 1054 being an electron-capture supernova (ECSN) \citep{Miyaji1980}, which involves the collapse of an oxygen-neon-magnesium core in an 8-10 M$_\odot$ progenitor \citep{Nomoto1982, Nomoto1987}.  This produces an explosion with a typical energy of 10$^{50}$ erg, compared to the canonical 10$^{51}$ erg from the collapse of an iron core.  However, low-energy explosion models, including ECSNe, are typically sub-luminous due to the low quantity of $^{56}$Ni synthesized during the explosion \citep{Kitaura2006}.  SN 2018zd has also been suggested to be an ECSN \citep{Zhang2020, Hiramatsu2021}, although further studies of both the Crab Nebula and SN 2018zd have favoured a low-mass core-collapse interpretation \citep{Callis2021, Temim2024}.
Some studies propose that the luminosity could be powered by shock interaction with circumstellar medium (CSM) \citep{Fesen1997, Sollerman2001, Smith2013}, although this is disfavoured by some models \citep{Hester2008, Yang2015} due to the required mass limiting the presence of freely expanding ejecta. Other studies propose that the central pulsar could have supplied the required energy \citep{Sollerman2001, Li2015}.

The discussion of CSM and pulsar-power draws parallels to another class of transients; superluminous supernovae (SLSNe).  Photometric observations are unable to distinguish between the power sources \citep[e.g.][]{Chen2023b}, and therefore, other information, such as nebular spectra \citep{Chevalier1992, Jerkstrand2017, Dessart2019, Omand2023}, polarization \citep{Inserra2016, Saito2020, Poidevin2022, Pursiainen2022, Poidevin2023, Pursiainen2023}, infrared emission \citep{Omand2019, Chen2021, Sun2022}, and radio emission \citep{Murase2015, Omand2018, Eftekhari2019, Law2019, Mondal2020, Eftekhari2021, Margutti2023} are used to try and diagnose the power sources of these supernovae. 
While we have access to extensive multiwavelength observations about the Crab, models of pulsar-driven supernovae generally do not make predictions out to 1000 years due to the extragalactic distances of those sources.  

The properties of the Crab pulsar and pulsar wind nebula (PWN) have been extensively studied.  The spin frequency and frequency derivative are 30 Hz and -4 $\times$ 10$^{-10}$ Hz s$^{-1}$ respectively~\citep{Staelin1968, Lyne1993, Lyne2015}.  The characteristic magnetic field of the pulsar is $\sim$ 8 $\times$ 10$^{12}$ G,  assuming $B_c = 6.4 \times 10^{19} \sqrt{P\dot{P}}$ G for pure magnetic dipole losses \citep{Kou2015}, and the current braking index $n$ is 2.51 $\pm$ 0.01 \citep{Lyne1993}.  Estimates of the initial pulsar spin period are usually in the range of 15 -- 20 ms \citep[e.g.][]{Kou2015}, but a study of the electron spectrum estimated a much faster initial spin of around 3 -- 5 ms \citep{Atoyan1999}.  A pulsar spinning at this rate could potentially supply the required energy to produce the luminosity observed in SN 1054.

In this work, we examine SN 1054 under the lens of the pulsar-driven supernova model to determine whether this scenario is consistent with the observed supernova and remnant properties and estimate the initial properties of the Crab pulsar and PWN.  We also examine the implications of a pulsar engine on the evolution of the pulsar and the supernova remnant.  In Section \ref{sec:modcon}, we overview the model and constraints from observations.  In Sections \ref{sec:res} and \ref{sec:imp}, we present the results from our analysis and discuss their implications.  Lastly, in Section \ref{sec:sum}, we summarize our findings.

\section{Model and Constraints} \label{sec:modcon}

\subsection{Model Overview}

The model we use is the generalized magnetar-driven supernova model first presented in \citet{Omand2024}, based on models of magnetar-driven kilonovae \citep{Yu2013, Metzger2019, Sarin2022}. We present a brief summary of the key components of the model here.

The spin-down luminosity of the pulsar is 
\begin{equation}
    L_{\rm SD}(t) = L_0 \left( 1 + \frac{t}{t_{\rm SD}} \right)^{\frac{1+n}{1-n}},
    \label{eqn:llasky}
\end{equation}
where $L_0$ is the initial spin-down luminosity, $t_{\rm SD}$ is the spin-down timescale, and $n$ is the braking index defined from $\dot{\Omega} \propto -\Omega^n$.  The total rotational energy is 
\begin{equation}
    E_{\rm rot} = \frac{n-1}{2}L_0t_{\rm SD} . 
    \label{eqn:elt}
\end{equation}

The evolution of the internal energy of the ejecta is
\begin{equation}
    \frac{dE_{\rm int}}{dt} = \xi (L_{\rm SD} + L_{\rm ra}) - L_{\rm bol} - \mathcal{P}\frac{d\mathcal{V}}{dt},
    \label{eqn:deintdt}
\end{equation}
where $L_{\rm ra}$ and $L_{\rm bol}$ are the radioactive power and emitted bolometric luminosity, respectively, $\mathcal{P}$ and $\mathcal{V}$ are the pressure and volume of the ejecta, and 
\begin{equation}
   \xi = 1 - e^{-At^{-2}},
   \label{eq:xi}
\end{equation}
is the fraction of spin-down luminosity injected into the ejecta \citep{Wang2015}, where
\begin{equation}
    A = \frac{3 \kappa_\gamma M_{\rm ej}}{4\pi v^2_{\rm ej}}
    \label{eq:leakage}
\end{equation}
is the leakage parameter and $\kappa_\gamma$ is the gamma-ray opacity of the ejecta.  We assume all energy from radioactive decay is emitted as gamma rays and that the ejecta has the same gamma-ray opacity for both radioactive heating and magnetar heating.

The ejected material accelerates with 
\begin{equation}
    \frac{dv_{\rm ej}}{dt} = \frac{ E_{\rm int}}{M_{\rm ej}R_{\rm ej}}.
    \label{eqn:dvdt2}
\end{equation}
due to the interaction with the pulsar wind nebula, and the supernova bolometric luminosity is 
\begin{align}
    L_{\rm bol} = & \frac{E_{\rm int}c}{\tau R_{\rm ej}} = \frac{E_{\rm int}t}{t_{\rm dif}^2} & (t \leq t_\tau), \label{eqn:lbol_pret} \\
    = & \frac{E_{\rm int}c}{R_{\rm ej}}, & (t > t_\tau),\label{eqn:lbol_postt}
\end{align}
where 
 \begin{equation}
     \tau = \frac{\kappa M_{\rm ej} R_{\rm ej}}{\mathcal{V}}
     \label{eqn:opdep}
 \end{equation}
is the optical depth of the ejecta, $\kappa$ is the optical ejecta opacity, 
\begin{equation}
    t_{\rm dif} = \left(\frac{\tau R_{\rm ej} t}{c}\right)^{1/2}
    \label{eqn:tdif}
\end{equation}
is the effective diffusion time, and $t_\tau$ is the time when $\tau = 1$.

The photospheric temperature is 

\begin{equation}
T_{\rm phot} (t) = 
\begin{cases}
    \left( \frac{L_{\rm bol}(t)}{4 \pi \sigma R_{\rm ej}^2}\right)^{1/4} & \text{ for } \left( \frac{L_{\rm bol}(t)}{4 \pi \sigma R_{\rm ej}^2}\right)^{1/4} > T_{\rm min}, \\
    T_{\rm min} & \text{ for } \left( \frac{L_{\rm bol}(t)}{4 \pi \sigma R_{\rm ej}^2}\right)^{1/4} \leq T_{\rm min}
\end{cases}
\label{eqn:tphot}
\end{equation}
where $T_{\rm min}$ is the temperature of the supernova after the photosphere begins to recede.

The constant opacity $\kappa$ is justified while the ejecta is ionised, but this opacity will drop once the photospheric temperature hits the ionization temperature ($\sim$ 6000 K) and the ejecta starts to recombine \citep{Popov1993, Dexter2013, Tsuna2024}.  Recombination effects stop being important once the diffusion time $t_{\rm dif}$ becomes much smaller than the dynamical time.  We do not include recombination effects in this model, which we note as a potential caveat of our fits.

\subsection{Priors and Constraints}

The historical observations of SN 1054 \citep{Clark1977, Collins1999} can provide two constraints on the luminosity of the supernova at various times.  The first is that the supernova was visible during the day for at least 23 days and the second is that the supernova was visible during the night for around 650 days \citep{Clark1977}.  After accounting for extinction \citep{Miller1973}, this gives apparent $V$-band magnitudes of roughly -4.5 and 6.0 respectively, with uncertainties of $\pm$ 0.5 -- 0.8 mag \citep{Collins1999}.  

The priors used in inference are determined by constraints from observations of the Crab Nebula and from previous modeling of different supernovae. The most constraining distance estimate to the Crab Nebula, as determined from very-long-baseline interferometry (VLBI) measurements of a giant pulse, is 1.90$^{+0.22}_{-0.18}$ kpc \citep{Lin2023}.  
The ejecta mass estimated from an optical study of neutral and ionized gas in the Crab Nebula is 4.6 $\pm$ 1.8 M$_\odot$ \citep{Fesen1997}, although the authors suggest that up to 4 M$_\odot$ could remain undetected.  Observations of absorption in the freely expanding ejecta outside the nebula suggest that component has $\gtrsim$ 1.7 M$_\odot$ \citep{Sollerman2000}, and later radiative transfer simulations of the gas and dust content of the Crab find 7.2 $\pm$ 0.5 M$_\odot$ of material should be present within all the ejected material \citep{Owen2015}.  Given these estimates, we conservatively set the limits of the prior to be between 3 and 9 M$_\odot$.  
The explosion energy, estimated from the 1500 km s$^{-1}$ velocity of the pulsar bubble \citep{Bietenholz1991}, must be much lower than the canonical 10$^{51}$ erg value for typical core-collapse supernovae; a value of 10$^{50}$ erg is typical from simulations of the collapse of low-mass stars \citep{Nomoto1982, Nomoto1987}, so we set the prior between 10$^{49}$ erg and 10$^{50}$ erg.  This assumes that the component of the ejecta outside the filaments does not carry significantly more kinetic energy that the filaments themselves.
The amount of $^{56}$Ni synthesized in these low energy explosions is generally not more than a few 0.01 M$_\odot$ \citep{Kitaura2006}, so we fix the nickel fraction (i.e., the fraction of the total ejecta that is nickel) to 0.005. However, we note that such small quantity of nickel does not significantly contribute to the light curve.

The expected spin-down time of the Crab pulsar is $\sim$ 30 years if the initial pulsar spin is $\sim$ 5 ms \citep{Atoyan1999} and the magnetic field stays constant over time, and larger if the pulsar is spinning slower.  Since this timescale is much larger than the timescale that the supernova was observed for ($<$ 2 years), the spin-down luminosity (Equation \ref{eqn:llasky}) should not evolve significantly over that time ($L_{\rm SD}(t) \approx L_0$ for $t \ll t_{\rm SD}$).  This means that we can not infer either the spin-down timescale or braking index (which may be significantly different from the currently measured value) unless the spin-down timescale is significantly smaller than previously thought.  A significantly smaller spin-down timescale would imply a higher magnetic field than currently inferred.  We set a prior between $10^{4}$ -- $10^{10}$ s for the spin-down time to determine if the spin-down time can be short, but keep $n$ fixed to 3 (the exact value of $n$ does not strongly affect our results, see Appendix \ref{sec:appcorner} for details) .  The prior on $L_0$ ranges from 10$^{39}$ erg s$^{-1}$ to 10$^{46}$ erg s$^{-1}$, spanning the range from where the pulsar has almost no effect on the supernova to where the pulsar luminosity is consistent with a superluminous supernova \citep{Omand2024}.  The current spin period of the Crab pulsar is 33 ms \citep{Lyne1993}, which gives the pulsar a current rotational energy of $\sim$ 2 $\times$ 10$^{49}$ erg for a 1.4 $M_\odot$, 12 km radius neutron star; which we use to motivate the  lower limit on the prior for $L_0$ and $t_{\rm SD}$.

The final quantities to infer are the optical and gamma-ray opacities, $\kappa$ and $\kappa_\gamma$ respectively, the plateau temperature $T_{\rm min}$, and the explosion time.  The optical opacity prior is set to 0.34 cm$^2$ g$^{-1}$, the typical value for a hydrogen-rich supernova \citep{Inserra2018}.  The prior on gamma-ray opacity is not well constrained, and so a wide prior of $10^{-4}$ to $10^4$ cm$^2$ g$^{-1}$ is used, although recent work suggests values of $\sim$ $10^{-2}$ -- 1 cm$^2$ g$^{-1}$ are suitable for synchrotron nebulae \citep{Vurm2021}.  The plateau temperature, which is the temperature of the ejecta when the photosphere starts to recede, could be significantly lower than the typical value of 6000 K from SLSNe \citep{Nicholl2017}, and we take a prior from 500 -- 10 000 K to reflect this.  It is worth noting that fixing the plateau temperature to 6000 K does not strongly affect our results (see Appendix \ref{sec:appcorner}). The unknown explosion time is sampled uniform prior of up to 300 days before the point where the supernova faded from the daytime sky.  All of the parameters and priors are summarized in Table \ref{tbl:prior}.

\begin{table*}
\centering
\begin{tabular}{ccccc}
   Parameter & Definition & Units & Prior/Value & Posterior Values \\ \hline
   $D$ & Distance & kpc & U[1.72, 2.12] & 1.92$^{+0.12}_{-0.13}$\\
   $L_0$ & Initial Magnetar Spin-Down Luminosity & erg s$^{-1}$ & L[$10^{39}$, $10^{46}$] & $L$(44.50$^{+0.97}_{-1.45}$)\\
   $t_{\rm SD}$ & Spin-Down Time & s & L[$10^{4}$, $10^{10}$] & $L$(5.73$^{+1.12}_{-0.86}$)\\
   $n$ & Magnetar Braking Index &  & 3 & \\
   $f_{\rm Ni}$ & Ejecta Nickel Mass Fraction &  & 0.005 & \\
   $M_{\rm ej}$ & Ejecta Mass & $M_{\odot}$ & U[3, 9] & 5.24$^{+2.09}_{-1.55}$\\
   $E_{\rm SN}$ & Supernova Explosion Energy & 10$^{49}$ erg & U[1, 10] & 5.53$^{+2.82}_{-2.80}$ \\
   $\kappa$ & Ejecta Optical Opacity & cm$^2$ g$^{-1}$ & 0.34 & \\
   $\kappa_\gamma$ & Ejecta Gamma-Ray Opacity & cm$^2$ g$^{-1}$ & L[$10^{-4}$, $10^{4}$] & $L$(-2.63$^{+3.52}_{-0.98}$) \\
   $T_{\rm min}$ & Photospheric Plateau Temperature & 1000 K & U[0.5, 10] & 2.82$^{+4.60}_{-1.23}$ \\
    &   Explosion Time & days & U[-300,-0.1] & -183.29$^{+105.12}_{-72.85}$ \\
\end{tabular}
\caption{The parameters and priors used in this study. Priors are either uniform (U) or log-uniform (L).  The values shown for the posterior are the mean and 1$\sigma$ uncertainties.  Posterior values denoted with $L$ are given in log-space.  The full posterior is shown in Appendix \ref{sec:appcorner}.}
\label{tbl:prior}
\end{table*}

\section{Results}  \label{sec:res}

We fit the historical observations of SN 1054 using the model and priors described in Section \ref{sec:modcon}. 
Inference is performed using the open-source software package {\sc{Redback}}~\citep{Sarin24_redback} with the {\sc{pymultinest}} sampler \citep{Buchner2014} implemented in {\sc{Bilby}} \citep{Ashton2019}. 
We sample in magnitude with a Gaussian likelihood. 
We constrain the priors for $L_0$ and $t_{\rm SD}$ such that the initial rotational energy of the pulsar is higher than 10$^{49}$ erg, which is a conservative estimate of the current pulsar rotational energy.

The light curve fit is shown in Figure \ref{fig:lcfit} and the posterior in Appendix \ref{sec:appcorner}.  The only parameters that matter for determining the initial total energy of the pulsar wind nebula are $L_0$, the initial spin-down luminosity, and $t_{\rm SD}$, the pulsar spin-down time.  The initial spin-down luminosity is most likely $\sim$ 10$^{43-45.5}$ erg s$^{-1}$, which is similar to the initial spin-down luminosity of pulsars that power SLSNe such as SN 2015bn \citep{Omand2024}.  This value is a factor $\gtrsim$ 10 more than the estimated initial radioactive luminosity, and radioactivity also decays off more quickly than the magnetar luminosity. Therefore, the magnetar will always be the dominant contributor to the supernova internal energy (Equation \ref{eqn:deintdt}).  The spin-down timescale is most likely around $1-100$ days, much lower than the expected 30 years for a fast-rotating pulsar with constant magnetic field \citep{Atoyan1999}.  
This likely implies that the magnetic field must have initially been much stronger than the current inferred field strength; we discuss this further in Section \ref{sec:pulsev}.

The fitted light curves show a broad distribution in both peak luminosity and explosion time due to the low number of constraining data.  It is unclear from historical constraints whether the supernova could have had a peak magnitude much brighter than -5 or an explosion time in the winter of 1054, since the first known records of a possible supernova sighting are in April \citep{Collins1999}.  The posterior for the explosion time does not show a strong correlation with any other parameters (Figure \ref{fig:corner}), while the distribution of peak luminosities shows slight correlations with $L_0$ and $t_{\rm SD}$.  If an upper limit were imposed on the peak luminosity, this would push the posterior towards higher $L_0$ and lower $t_{\rm SD}$, in agreement with the general behaviour found in \citet{Omand2024}, and imply an even higher initial poloidal magnetic field.  None of our results or their implications would be significantly affected by peak luminosity or explosion time constraints.

\begin{figure}
\includegraphics[width=0.9\linewidth]{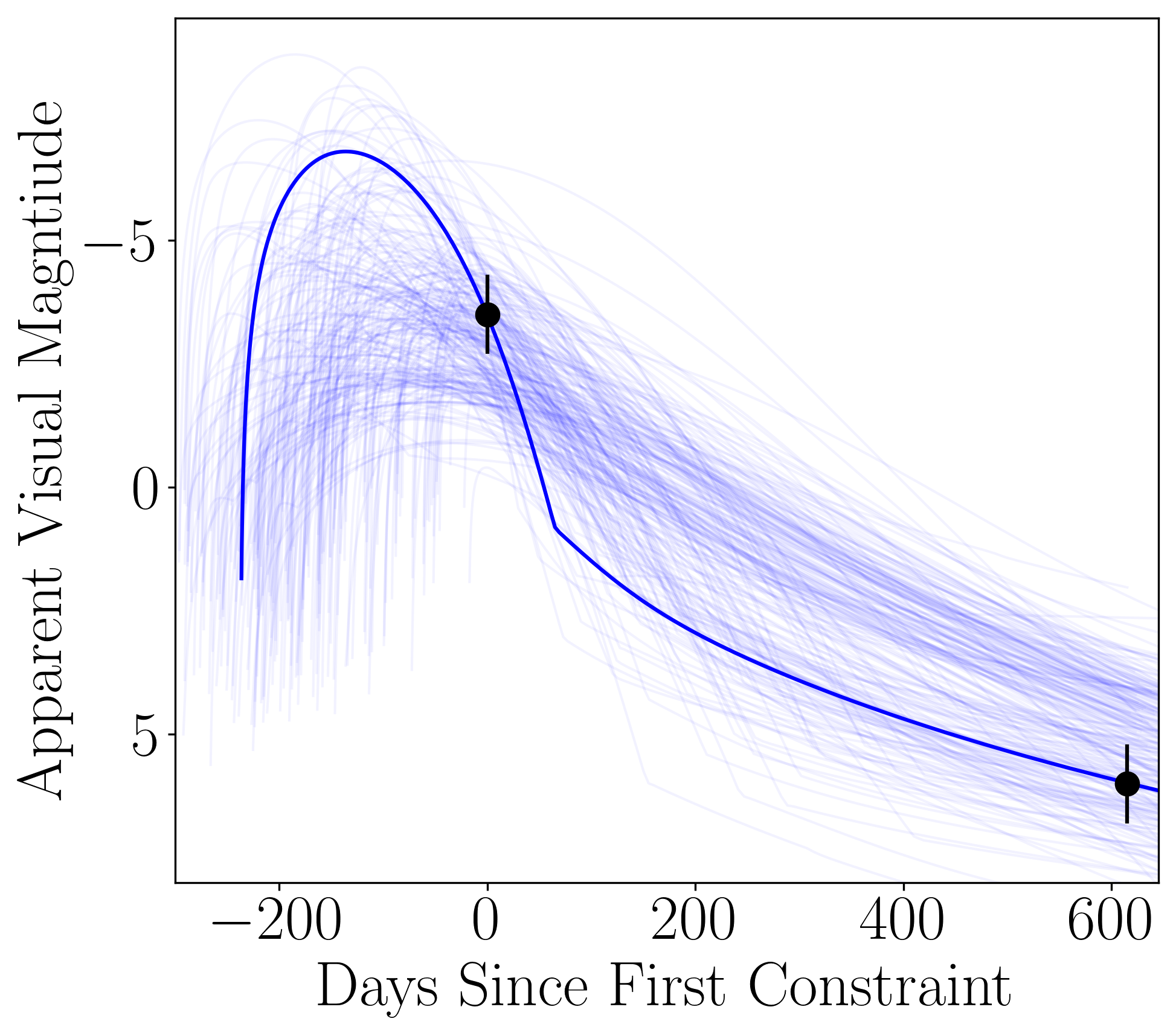}
\caption{Fitted dust-corrected light curves for SN 1054.  The blue lines indicate 300 models drawn randomly from the posterior, and the thick blue line indicates the most likely model.  Since we sample over distance, the distance modulus ($m_V - M_V$) varies between 11.2 and 11.6.}%
\label{fig:lcfit}
\end{figure}

The ratio of the calculated diffusion time from the model at the time of the first observation to the dynamical time of the first observation is $<$ 1/2 for 68\% of the posterior and $<$ 1/10 for 36\% of the posterior.  About 18\% of the posterior has $T_{\rm phot} > 6000$ K at the time of the first observation.  Assuming $t_{\rm dyn}/t_{\rm dif} > 2$ is a reasonable criterion for when recombination effects become unimportant, we find that only $\sim$ 32\% of the posterior would be affected by recombination at the first observation epoch.  Assuming recombination is important only when both $t_{\rm dyn}/t_{\rm dif} < 2$ and $T_{\rm phot} < 6000$ K, we still find that $\sim$ 32\% of the posterior would be affected by recombination at the first observation epoch, since all of the samples with high diffusion time also have low photospheric temperatures.  None of the supernovae from our posterior are affected by recombination at the second epoch.

\begin{figure}
\includegraphics[width=0.9\linewidth]{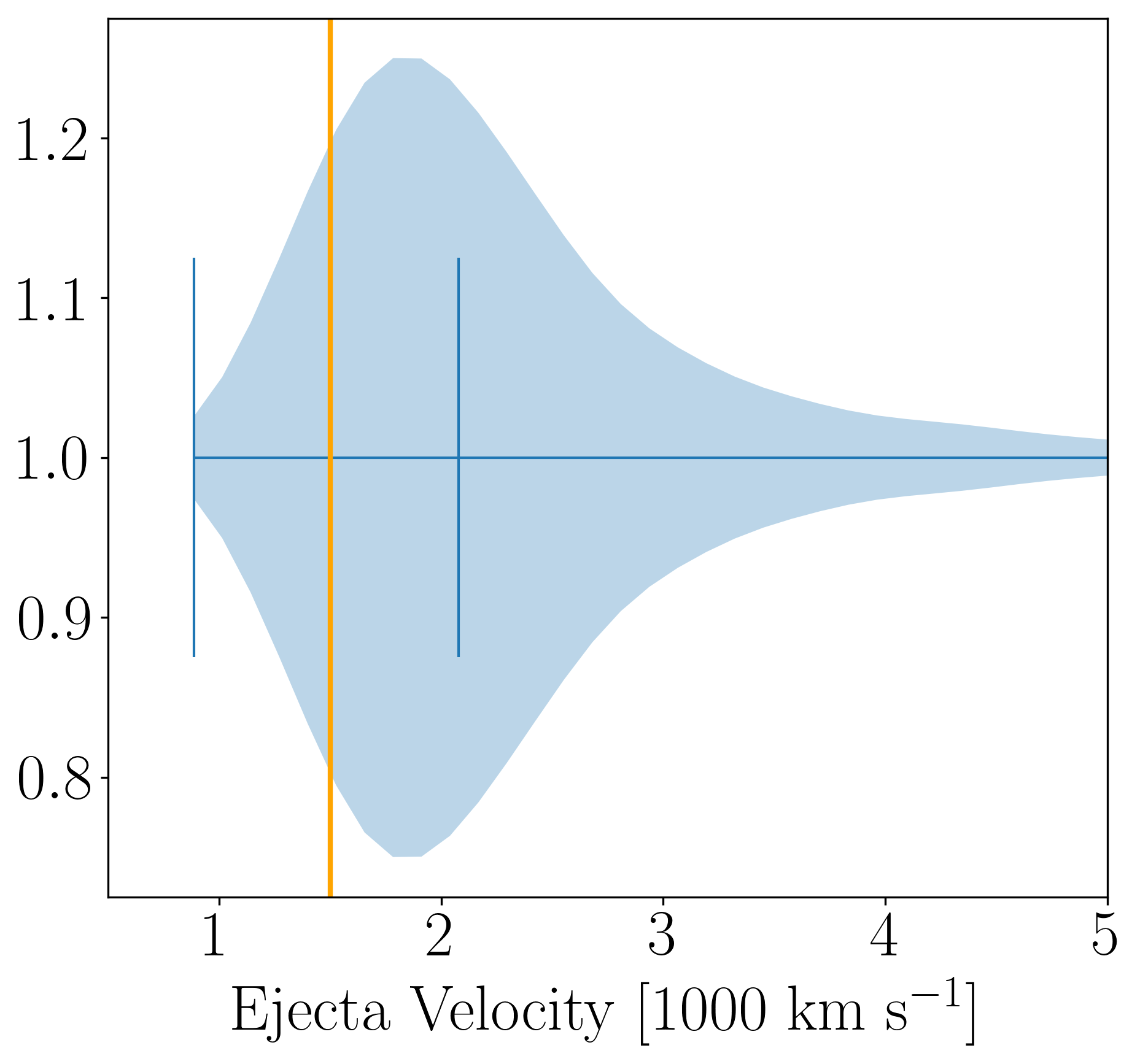} \\
\includegraphics[width=0.9\linewidth]{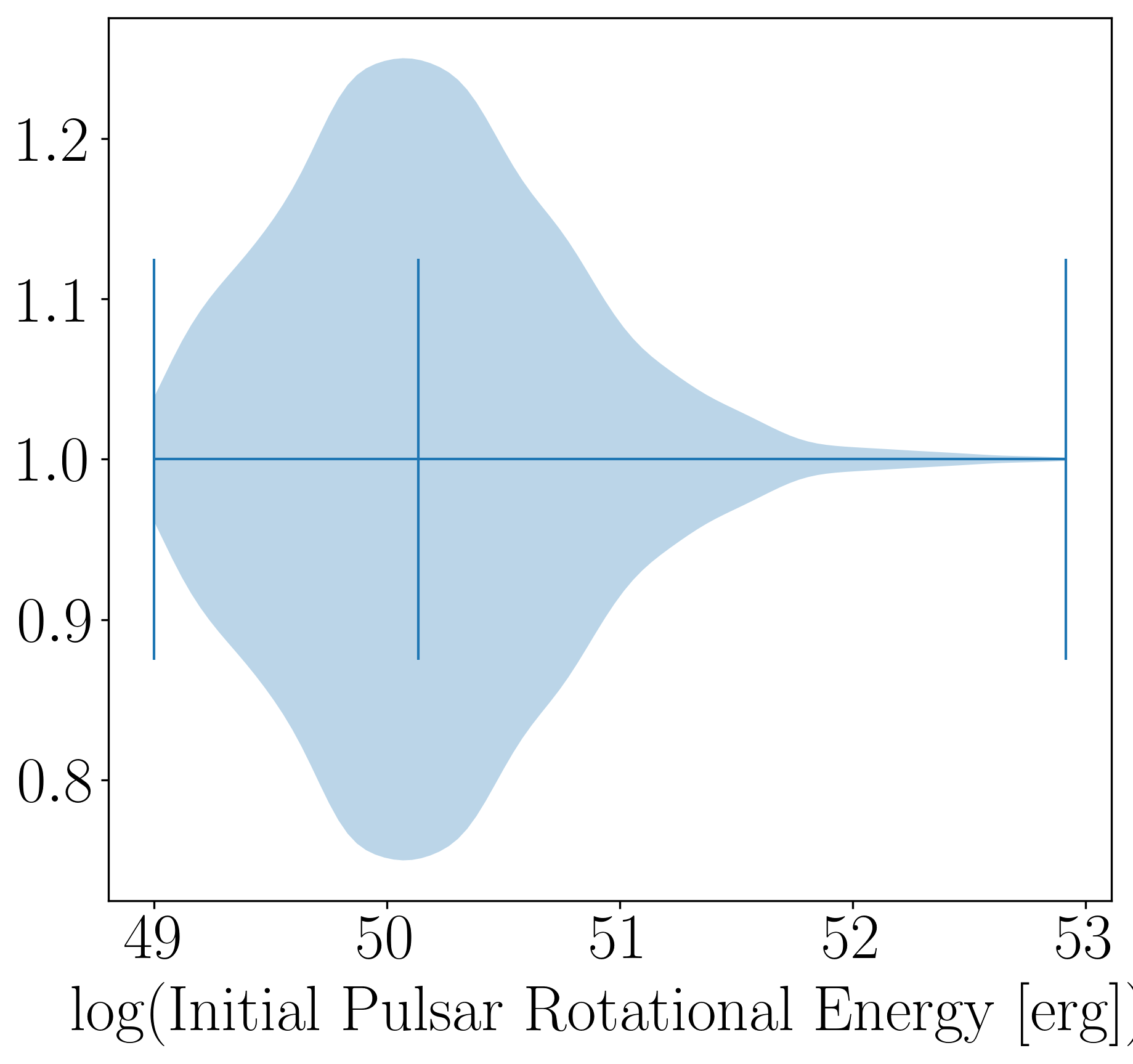} \\
\caption{(Top) The distribution of final ejecta velocities inferred for SN 1054, with the value of the Crab Nebula forward shock \citep{Bietenholz1991} shown with an orange line.  (Bottom)  The distribution of initial pulsar spin-down energies inferred for SN 1054.  The outer bounds and median values for each distribution are shown with blue lines.}%
\label{fig:vdist}
\end{figure}

The posterior distributions of the final ejecta velocity and initial pulsar rotational energy, assuming vacuum dipole spin-down, are shown in Figure~\ref{fig:vdist}.  The ejecta velocity is defined here as the velocity where $E_{\rm kin} = \frac{1}{2}M_{\rm ej}v_{\rm ej}^2$; this is sometimes known as the scaling velocity or bulk ejecta velocity.  The median values of the two distributions are 2000 km s$^{-1}$ and 1.4 $\times$ 10$^{50}$ erg respectively.  Most of the inferred ejecta velocities are only slightly higher than the measured value of 1500 km s$^{-1}$ of the forward shock \citep{Bietenholz1991}, although only 14$\%$ of the distribution is below that value.  The initial rotational energy peaks at only slightly higher than the maximum value from the explosion energy prior of 10$^{50}$ erg, and is similar to the values inferred for the SN Ic-BL SN 2007ru and USSN iPTF14gqr and lower than those inferred for the SLSN SN 2015bn and FBOT ZTF20acigmel \citep{Omand2024}.  Using scaling relations for a  1.4 $M_\odot$, 12 km neutron star 

\begin{align}
    E_{\rm rot} = & \, 2.6 \times 10^{52} P_{0, -3}^{-2} \text{ erg}\label{eqn:erotscale}, \\
    L_0 = & \, 2.0 \times 10^{47} P_{0, -3}^{-4} B_{14}^2 \text{ erg s$^{-1}$} \label{eqn:l0scale}, \\
    t_{\rm SD}= & \, 1.3 \times 10^5 P_{0, -3}^2 B_{14}^{-2} \text{ s},  \label{eqn:tsdscale}
\end{align}
to convert this energy into an initial spin period gives $P_0 = 16.3\pm10.5$ ms with a median value of 13.8 ms, which is consistent with both the values of 15 -- 20 ms derived from extrapolating backwards from current conditions (See Appendix \ref{sec:appspin}) and the value of 5 ms estimated from the radio spectrum of the pulsar wind nebula \citep{Atoyan1999}.

\begin{figure}
\includegraphics[width=0.89\linewidth]{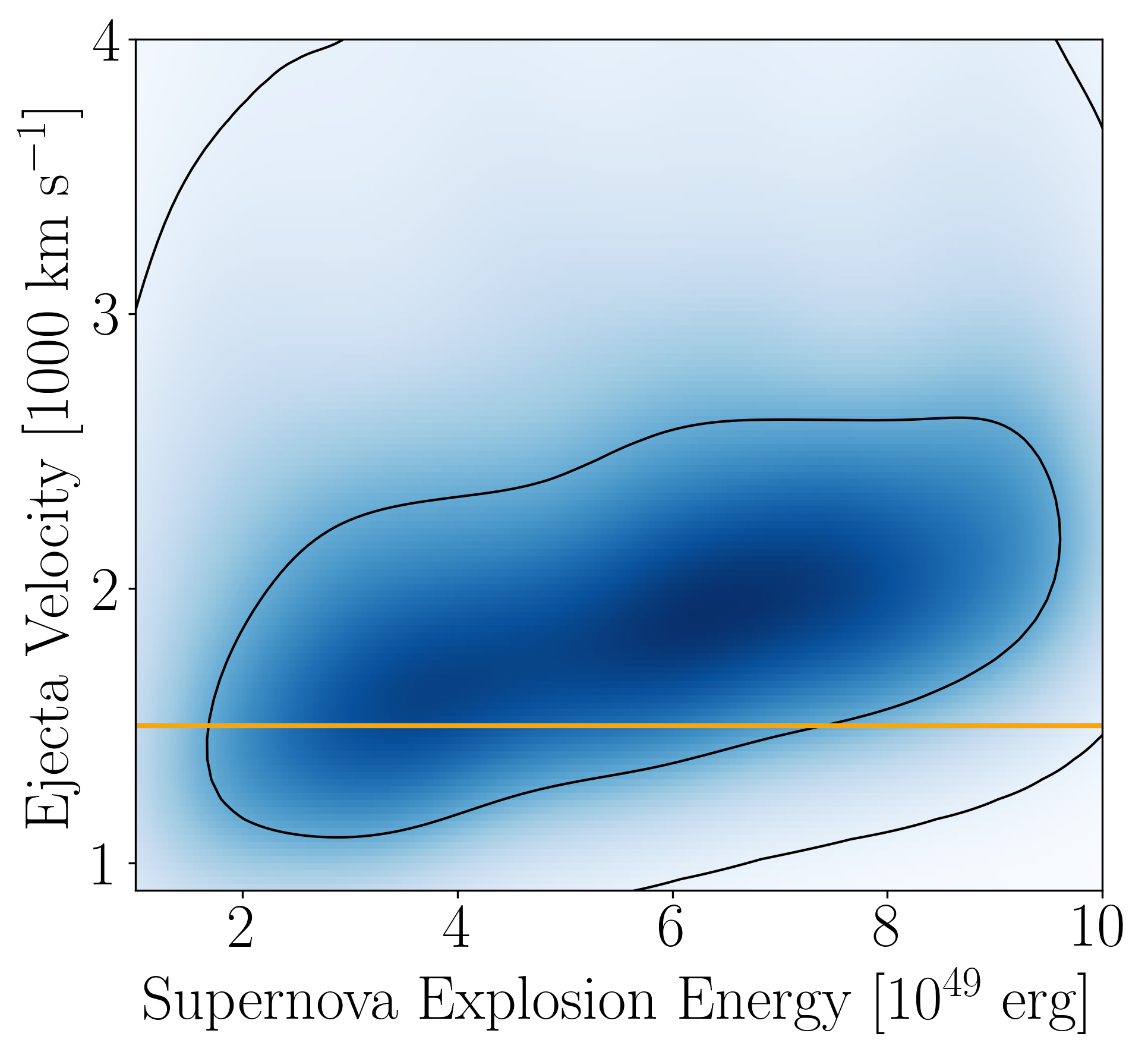} \\
\includegraphics[width=0.89\linewidth]{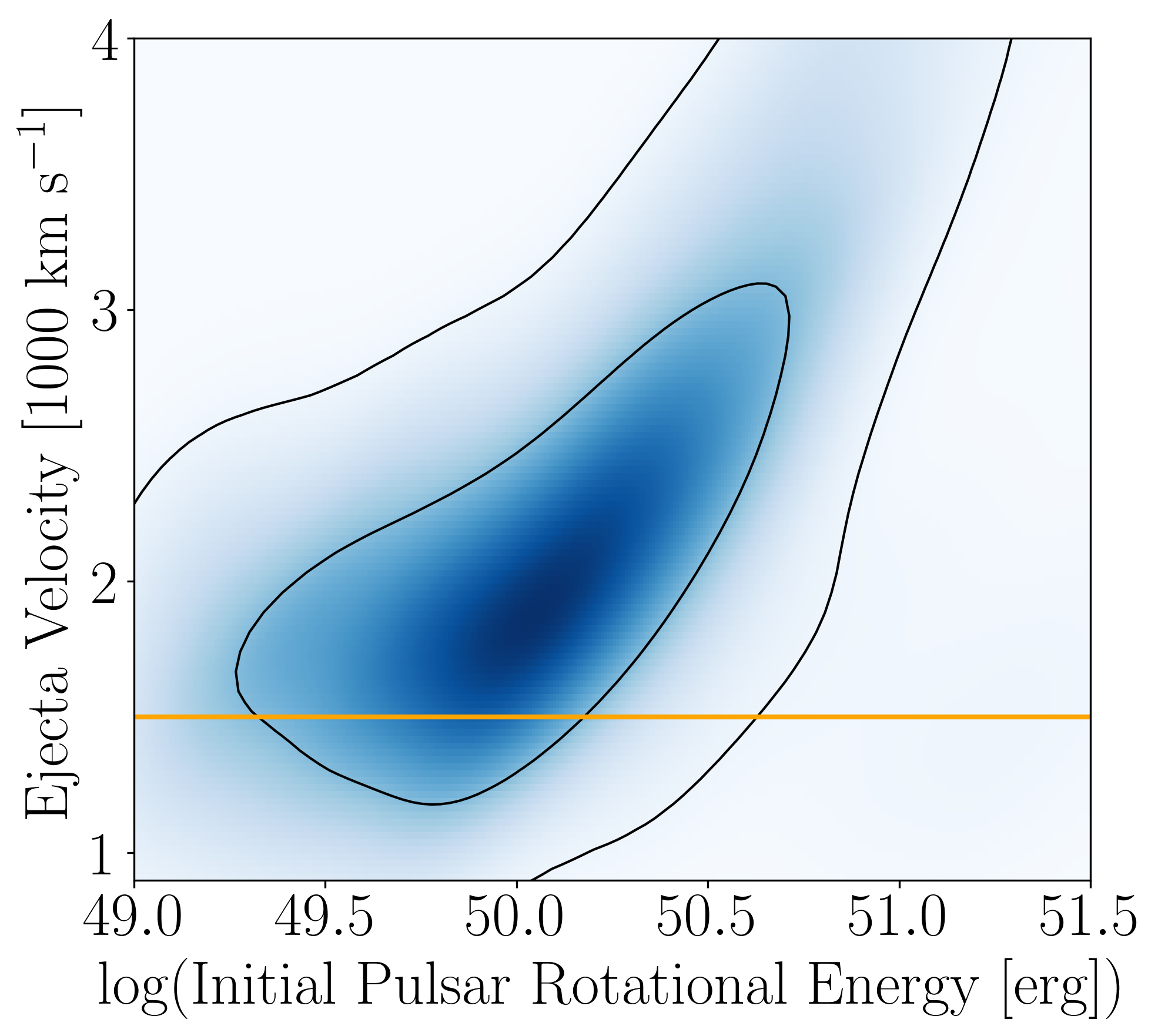} \\
\includegraphics[width=0.89\linewidth]{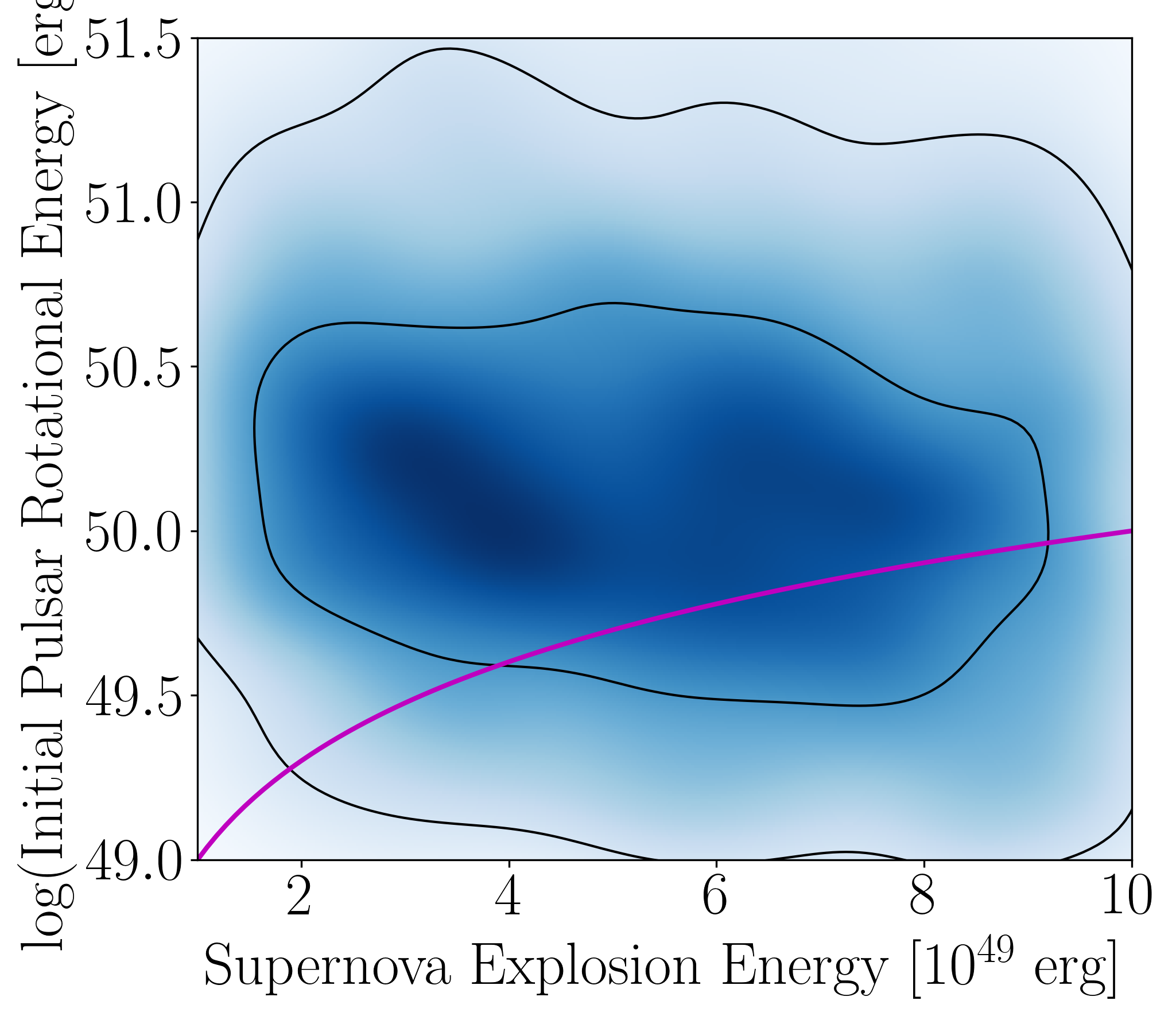}
\caption{The two-dimensional posterior distributions of ejecta velocity $v_{\rm ej}$, supernova explosion energy $E_{\rm SN}$, and initial pulsar rotational energy $E_{\rm rot}$.  The black contours encompass the $50\%$ and $90\%$ credible intervals.  The orange lines in the top and middle figures show the velocity of the Crab Nebula forward shock \citep{Bietenholz1991}, and the magenta line in the bottom figure shows where $E_{\rm rot} = E_{\rm SN}$.  Everything above the magneta line is expected to exhibit blowout \citep{Blondin2017}.}%
\label{fig:esn_erot_vej}
\end{figure}

The two-dimensional posterior distributions of ejecta velocity $v_{\rm ej}$, supernova explosion energy $E_{\rm SN}$, and initial pulsar rotational energy $E_{\rm rot}$ are shown in Figure \ref{fig:esn_erot_vej}.  Both $E_{\rm SN}$ and $E_{\rm rot}$ show weak correlations with $v_{\rm ej}$, while the two energies are not strongly correlated with each other.  Most models with explosion energies close to 10$^{50}$ erg show velocities higher than 1500 km s$^{-1}$, justifying the upper limit of the explosion energy prior.  If the posteriors were constrained to have velocities lower than this limit, the explosion energy would likely have $E_{\rm SN} \lesssim 4 \times 10^{49}$ erg, while the rotational energy would roughly lie between 5 -- 10 $\times$ $10^{49}$ erg, giving a spin period of 16 -- 22 ms.  It is worth noting that the supernova explosion energy is not well constrained on its own, and only correlates with the ejecta velocity.  Thus, our model can not shed light into the explosion mechanism or distinguish between electron capture and iron-core collapse explosions.  Examining the correlation in the energies shows that most of the posterior, 75$\%$, has $E_{\rm rot} > E_{\rm SN}$, and this percentage will rise when selecting for lower velocity models.  Supernovae with $E_{\rm rot} > E_{\rm SN}$ undergo blowout, where the PWN forward shock can expand past the inner region of the ejecta, changing the structure of the ejecta and remnant; we discuss this further in Section \ref{sec:snrev}.

\section{Implications}  \label{sec:imp}

\begin{figure*}
    \centering
    \includegraphics[width=\textwidth]{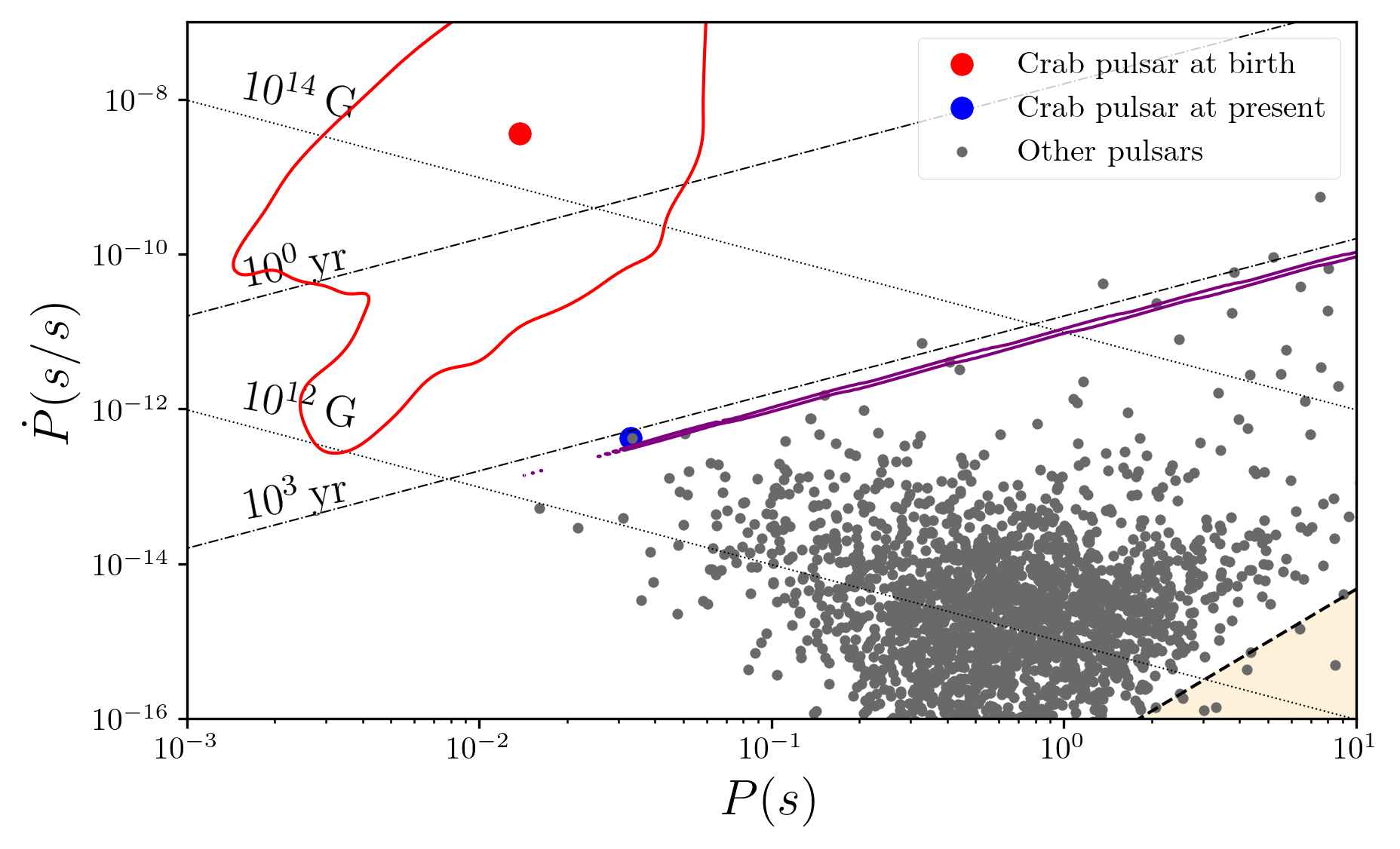}
    \caption{Period and Period derivative $P\dot{P}$ diagram showing the inferred birth and present locations of the Crab pulsar in red and blue respectively, and other pulsars in the ATNF catalog gathered via the {\sc psrqpy} package~\citep{psrqpy}. The solid red line indicates the $95\%$ credible interval of the posterior density distribution.  The purple dots indicate the 95$\%$ credible interval of the predicted present day locations of the posterior after undergoing vacuum dipole spin-down and Ohmic dissipation.  We also show lines of characteristic ages, constant magnetic fields, and the region below the pulsar `death line' indicated in yellow.}
    \label{fig:ppdot}
\end{figure*}

\subsection{Evolution of the Pulsar} \label{sec:pulsev}
The simple theory for the evolution of neutron stars is that they are born with rapid spins, and spin down through vacuum dipole radiation with a constant magnetic field~\citep{Pacini1967, borghese_2020}. On a period-period derivative ($P\dot{P}$) diagram, this translates into neutron stars being on on the top left of the $P\dot{P}$ diagram and decaying down towards the bottom right on lines of constant magnetic fields.  

In Figure~\ref{fig:ppdot}, we show a $P\dot{P}$ diagram with the current location of the Crab pulsar (in blue), the inferred location of the pulsar at birth (solid red for the mean of the posterior, with the contour encompassing the $95\%$ credible interval), and the locations of other pulsars obtained from the ATNF catalog~\citep{Manchester2005} via the package {\sc psrqpy}~\citep{psrqpy} in gray. 
We also show lines of characteristic ages, constant magnetic fields, and the region below the pulsar `death line' indicated in yellow.
Given the inferred location at birth of the Crab pulsar, the canonical model for neutron star evolution would predict that it evolves along a diagonal line; spinning down but remaining above the $\unit[10^{14}]{G}$ magnetic-field line and would be more consistent with the location at present day of other ``magnetars'' rather than its true current location around other pulsars. 
This would seem alarming but the simple theory described above is known to be incorrect in a multitude of ways. 
For example, neutron star magnetic fields are expected to decay due to ohmic dissipation~\citep{Igoshev+21}, and neutron stars, especially newly born neutron stars, are expected to spin-down through mechanisms other than just vacuum dipole radiation~\citep{Melatos1999, Lasky2017, Sarin2018, Sarin2020b}. 
Another theory, motivated by detailed magnetohydrodynamic simulations suggests that neutron stars are not in fact born with high poloidal (external) magnetic fields but rather small-scale turbulent magnetic fields that later grow to large-scales via an inverse cascade~\citep{Sarin2023_cascades}. 
We note that the latter would be at odds with the model we used to fit the historic supernova observations. Given we have, in theory, the location of the Crab pulsar at two evolutionary stages, it is tempting to attempt to interpret how the Crab pulsar must have evolved. We of course, do emphasize that the inferred posterior on the birth location is broad, as expected, such that a simple vacuum dipole radiation model with no evolution of the large-scale magnetic field, need not be ruled out. 

The disparity in terms of magnetic fields suggested by the birth and present day location, immediately suggests that the magnetic field must have decayed over the $\approx \unit[1000]{yrs}$ since the supernova. 
The decay of the large-scale, poloidal magnetic field under ohmic dissipation is expected to follow~\citep{Pons2007, Sarin2023_cascades}
\begin{equation}
B_{\rm p} = \frac{B_{\rm p,0}}{(1 + t/\tau)^{\alpha}} ,
\end{equation}
where the exponent, $\alpha \sim 1.3$, and $\tau = 800$ yr is the timescale when the magnetic-field starts to decay. 
Assuming this magnetic-field evolution, and for simplicity, vacuum-dipole radiation, we evolve the inferred birth location posteriors forward in time till present day. These projected present day locations are shown in purple (as a $95\%$ credible interval region). 
The projected locations of the crab pulsar on the $P\dot{P}$ diagram suggest two modes, one that would place the crab pulsar more in line with galactic magnetars, and another that would be more consistent with the true, present-day location of the crab pulsar. The former mode can be dismissed entirely (at least under the assumption of vacuum dipole spin down and Ohmic dissipation with the above parameters). However, the consistency of the latter mode is tantalising, suggesting some dissipation of the poloidal magnetic field of the crab pulsar beginning in the last $\approx 200$ yrs. We note that the values chosen for the decay timescale and exponent above are inconsistent with expectations of Ohmic dissipation from simulations~\citep{Pons2007}. However, the general decay behaviour could be recreated through other scenarios such as fall-back accretion. Moreover, inclusion of other more complex spin-down mechanisms, such as gravitational waves could further reconcile the differences from the magnetic-field decay behaviour compared to numerical simulations.  

Several mechanisms for magnetic field amplification have been suggested to explain the origin of magnetar-strength magnetic fields.  Field amplification from convective dynamos \citep{Raynaud2020}, magnetorotational instability-driven turbulent dynamos \citep{Reboul-Salze2021, Guilet2022}, and $\alpha\Omega$ dynamos \citep{Reboul-Salze2022} can amplify the dipole component of the magnetic field up to $\sim$ 10$^{15}$ G after the proto-neutron star contracts for an initial period $\sim$ 1 ms.  However, these amplification mechanisms scale down as the neutron star spin period increases, meaning they likely could not reproduce the birth properties of the Crab pulsar over a majority of the $P\dot{P}$ posterior.  Supernova fallback can also trigger a Tayler-Spruit dynamo in slower rotating proto-neutron stars \citep{Barrere2022, Barrere2023}.  Depending on how the field saturates, dipole fields of $\sim$ 10$^{15}$ may require rotation periods of $\lesssim$ 5 ms \citep{Spruit2002} or $\lesssim$ 25 ms \citep{Fuller2019}, which make it a viable amplification mechanism for the Crab in the latter case.


\subsection{Evolution of the Supernova Remnant} \label{sec:snrev}

The low inferred explosion energies imply that the initial velocity of the ejecta was only a few hundred km s$^{-1}$, and that a majority of the kinetic energy of the current Crab ejecta is due to the acceleration of the ejecta by the pulsar wind nebula.
In supernovae where the pulsar can deposit energy in excess of the initial supernova energy into the ejecta, the ram pressure of the ejecta can not confine the pulsar bubble, leading the pulsar bubble to break out through the shell \citep{Blondin2017, Suzuki2017}. 
This is the case over most of the posterior (see Figure \ref{fig:esn_erot_vej}), but considering that some of the PWN escapes the system without interacting with the ejecta, the PWN energy that couples to the ejecta may be more comparable to the explosion energy. 

The injected pulsar energy will cause the ejecta shell to become Rayleigh-Taylor unstable, leading to the formation of a filamentary structure similar to what is observed \citep{Jun1998, Bucciantini2004, Porth2014}.   The Rayleigh-Taylor instabilities create pressure waves that can deform, but not disrupt, the termination shock front \citep{Camus2009, Porth2014}; this may cause asymmetry in the photons emitted by the pulsar wind nebula but will not affect the large scale structure of the remnant \citep{Blondin2017}.  

Simulations from \citet{Blondin2017} show that once the pulsar wind nebula forward shock moves from the inner ejecta, with a flat density profile, to the outer ejecta, with a steep density profile ($\rho \approxprop r^{-9}$), the shock is strongly accelerated compared to the ejecta (see their Figure 5), leaving the most massive filaments behind.  Given that the observed shock velocity is about a factor $\sim$ 2 greater than the velocity of the innermost filaments \citep{Clark1983}, Figure 5 from \citet{Blondin2017} implies that the time when blowout started to occur must have been around 50-200 years post-explosion.  However, they use a constant spin-down luminosity instead of one that decreases with time, so this is only an upper limit.

For a constant pulsar spin-down, with $E_{\rm inj} = L_0t$, we can put an independent, although speculative, constraint on the initial spin period of the pulsar.  Using Equation 6 from \citet{Blondin2017} for the blowout time $t_{\rm tr}$, we find that $E_{\rm inj}/E_{\rm SN} = 1.5t/t_{\rm tr}$.  Since $t/t_{\rm tr} \sim 10$ gives the value of the shock velocity consistent with the Crab, this implies that $E_{\rm inj}/E_{\rm SN} \sim 15$.  If we assume that this is true regardless of the time dependence of the pulsar luminosity, that most of the rotational energy has already been emitted, and that the explosion energy is $10^{49-50}$ erg, then the initial rotational energy of the pulsar should be $\sim 1.5 \times 10^{50-51}$ erg.  This is consistent with our Figures \ref{fig:vdist} and \ref{fig:esn_erot_vej}, but does exclude many of the slower rotating pulsars and a few of the fastest.  The spin periods for these pulsars would be 4 -- 13 ms.

This scenario also implies that the freely expanding ejecta outside the filaments can not carry a significant amount of kinetic energy, and must therefore have a density profile that falls off more rapidly than $r^{-4}$ \citep{Sollerman2000}.  Further observations of the inner region of the Crab Nebula with sensitive, high-resolution instruments such as the James Webb Space Telescope (JWST) may help elucidate the low-velocity filament structure and the status of blowout within the nebula, placing further constraints on the energy oinjection of the Crab pulsar over the first few centuries of its lifetime.

The Crab is expanding within a low density void in the HI distribution \citep{Romani1990, Wallace1994, Wallace1999}.  The low ISM density means that the supernova forward and reverse shocks should be faint, which is consistent with their current non-detections \citep{Mauche1989, Predehl1995, Frail1995, Seward2006}.  Due to the low inferred explosion energy, the supernova forward and reverse shocks should have velocities not significantly higher than the 2500 km s$^{-1}$ inferred from \citet{Sollerman2000}.

\subsection{Light Echoes}

The light associated with the luminous peak of the supernova can scatter off of dust clouds around the remnant, which can be detected after a time delay.  These light echoes have been detected for several historical supernovae \citep{Crotts1989, Sugerman2006, Rest2005, Rest2008a}, and both Tycho's SN \citep{Krause2008, Rest2008a} and Cas A \citep{Krause2008b, Rest2008a, Rest2011a, Rest2011b} were able to be classified as a Type Ia and Type IIb SN respectively because of light echo spectroscopy.  Despite its old age and low-density environment, it may be possible to detect light echoes from SN 1054 as well.

Detection of light echoes from SN 1054 could provide a direct test of the power source of the supernova.  The brightness evolution of the light echo would provide a better sampled light curve than the historical observations.  The presence or absence of a plateau would provide a diagnostic of whether the supernova was a Type II-P/IIn-P \citep{Smith2013} or something else, and better time resolution around the supernova peak would provide stronger constraints on the initial pulsar properties.  

The spectrum in the early phase would also show slightly broader lines than the currently inferred filament and shock velocity due to the photosphere receding from the expanding envelope.  These lines would likely have velocities around 2500 km s$^{-1}$, similar to that inferred by C IV absorption \citep{Sollerman2000}.  Due to the slow, high-opacity ejecta, the transition to the nebular phase would likely take several years, so the narrowing of the lines as the photosphere recedes would likely not be detectable.  The early spectrum would likely resemble an SLSN-II without narrow features \citep{Kangas2022}, showing broad Balmer emission lines, sometimes with a P Cygni profile, as well as absorption lines from Na I, He I, Fe II, Sc II and emission from Mg I] and Ca II.  The spectrum may also develop H$\alpha$ and H$\beta$ emission lines a few weeks after maximum light.  The supernova would not be expected to show narrow hydrogen lines in this scenario, in contrast to a Type IIn or IIn-P.

\subsection{Comparison with Previous Works}

Several works have suggested that the Crab pulsar could have contributed to the luminosity of SN 1054 \citep{Schramm1977, Chevalier1992, Sollerman2001}, but note that the pulsar wind nebula would not be a significant source of supernova luminosity unless the nebula luminosity was orders of magnitude higher than it currently is.  This would happen if either the pulsar was rapidly rotating \citep{Atoyan1999} or the magnetic field was much higher than it currently is, as we find.

SN 1054 was previously fit with a pulsar-driven model by \citet{Li2015}, although their results and methodology are significantly different than ours.  The most obvious difference is that they do not use a Bayesian inference code to do their fit, and thus can not show parameter posteriors or show the uncertainty or correlation on their inferred values.  They also used fixed values for several parameters, including gamma-ray opacity, ejecta mass, explosion energy, and distance, instead of marginalizing over them.  They assume an electron scattering opacity of $\kappa = 0.2$ cm$^2$ g$^{-1}$ instead of $\kappa = 0.34$ cm$^2$ g$^{-1}$, which is the standard value for hydrogen rich supernovae.  They also fix a temperature at peak and assume a bolometric correction for both epochs instead of self-consistently calculating what the observed emission would be in the human visual band.

The resulting spin periods measured by \citet{Li2015} are smaller than what we infer, although the spin-down timescales are similar, implying a smaller magnetic field.  The rotational energies inferred by their fits are $5-20$ $\times$ 10$^{50}$ erg, which are higher than our median value.  Our inferred initial pulsar luminosity range is an order of magnitude lower, and their highest pulsar luminosity is similar to that inferred for an FBOT or BL-Ic SN \citep{Omand2024}.  This extra energy causes their ejecta to expand much more rapidly than inferred either by our models or by observations.

An alternate scenario for explaining the properties of the Crab supernova and remnant is interaction with dense CSM ejected prior to the supernova, as detailed in \citet{Smith2013}.  While an analysis of the pulsar + CSM scenario is beyond the scope of this work, and would likely not be useful due to a lack of observational constraints, it is worth noting how interaction would affect our inferred supernova and pulsar parameters.  Since CSM interaction converts kinetic energy into radiated energy, the parameters would be consistent with a less luminous supernova with faster ejecta.  There are two ways to achieve this, with vastly different implications implications on the evolution of the pulsar.  One is that the magnetic field can increase even further, decreasing the spin-down time and supernova luminosity while increasing the ejecta velocity  \citep{Omand2024}.  The second is that the rotational energy can decrease, decreasing the supernova luminosity and ejecta velocity, with a corresponding increase to the supernova explosion energy to maintain or increase the velocity.

\subsection{Comparison to Other Objects}

The broad inferred spin period and magnetic field distributions and lack of many observational constraint make it difficult to determine what exactly the modern analogue of SN 1054 is.  In particular, not having a strong constraint on the peak luminosity allows for possible analogues to range from normal Type II SNe, to luminous SNe (LSNe), to SLSNe, although we note that the boundaries between these classes are not well defined and there may simply be one continuous luminosity distribution.

SLSNe can show peak absolute magnitudes as faint as around -20 \citep{Kangas2022, Chen2023a, Gomez2024}, which is consistent with the most luminous light curves from the SN 1054 posterior sample.  Sample studies of SLSNe-I tend to show spin periods $\lesssim$ 8 ms and dipole magnetic fields of $\sim$ 1 -- 5 $\times$ 10$^{14}$ G for spin periods $\gtrsim$ 4 ms, and $\sim$ 0.1 -- 5 $\times$ 10$^{14}$ G for spin periods $\lesssim$ 4 ms \citep{Nicholl2017, Chen2023b}.  A sample study of SLSNe-II found similar parameters with spin periods $\lesssim$ 5 ms and dipole magnetic fields of $\sim$ 0.5 -- 10 $\times$ 10$^{14}$ G \citep{Kangas2022}.  The majority of ejecta masses for both types of SLSNe are inferred to be between 3 -- 10 $M_\odot$, similar to the possible range for SN 1054.  These parameters are consistent with part of the distribution for SN 1054, although not with the mean inferred value, which has a similar magnetic field but slower spin period.  

LSNe show peak absolute magnitudes between $\sim$ -18 and -20 \citep{Gomez2022, Pessi2023}, which is more consistent with the majority of  the SN 1054 posterior than SLSNe.  The small sample of LSNe-II has no estimated pulsar parameters, and \citet{Pessi2023} prefer a CSM power source because of various other observational constraints.  \citet{Gomez2022} present a sample of LSNe-I, and find the majority of them to have a significant contribution from a magnetar engine.  The spin period and magnetic field distribution they find is broad, but does have several SNe with high magnetic field and slow spin period, similar to the inferred median initial values of the Crab pulsar.  \citet{Rodriguez2024} also claim that most stripped-envelope supernovae show signs of central engine activity at late times.

The progenitors of LSNe-I and SLSNe-I can vary greatly in mass due to the differing amount of material that can be stripped from the star before the explosion \citep{Blanchard2020, Gomez2022}.  These SNe tend to be found is low-mass, star-forming galaxies \citep{Lunnan2014, Leloudas2015b, Angus2016, Schulze2018, Orum2020}, and are typically thought to come from progenitors with zero age main sequence (ZAMS) masses $\gtrsim$ 18 $M_\odot$ \citep{Chen2023b}, much larger than expected for SN 1054.  For LSNe-II and SLSNe-II, the progenitors are expected to be less massive red supergiant (RSG) or yellow supergiant (YSG) stars \citep{Kangas2022, Pessi2023}, which is more consistent with the mass and composition expected for the progenitor of SN 1054.

\section{Summary}  \label{sec:sum}

We use a model for a pulsar-driven supernova \citep{Omand2024} to compare with historical and contemporary observations and constraints on the Crab supernova.  We perform the fit using the Bayesian open-source software {\sc{Redback}}~\citep{Sarin24_redback} and find that the most likely value for the initial spin-down luminosity is $\sim$ 10$^{43-45.5}$ erg s$^{-1}$ and for the initial pulsar spin-down timescale is around 1 -- 100 days.  These imply an initial rotational energy of $\sim$ 10$^{50}$ erg and an initial spin period of $\sim$ 14 ms.  These also imply an initial magnetic field of $\sim$ 10$^{14-15}$ G, which is orders of magnitude higher than the current characteristic magnetic field.  The inferred bulk ejecta velocities are around 2000 km s$^{-1}$, which is similar to the current observed velocities of the PWN forward shock and filaments.

The large initial field implies that the magnetic field must have decayed over the lifetime of the pulsar.  Ohmic dissipation along with vacuum dipole spin-down may be able to reproduce the inferred evolution, but may also require other spin-down and field dissipation mechanisms.  
The high initial rotational energy compared to the explosion energy means that the supernova probably underwent pulsar bubble blowout, which causes the PWN forward shock to accelerate and leave behind the material in the filaments.  The slow PWN shock velocity gives an independent, although speculative, spin period constraint of 4 -- 13 ms.
The pulsar-driven scenario could be tested and constrained with light echo photometry and spectroscopy, particularly around the supernova peak.
SN 1054 shares similarities with both hydrogen-rich and hydrogen-poor LSNe and SLSNe, giving it a wide range of possible modern analogues.

\section*{Acknowledgements}

The authors thank the anonymous referee for their helpful comments.  The authors also thank Jesper Sollerman for his helpful discussions. 
C. M. B. O acknowledges support from the Royal Society (grant Nos. DHF-R1-221175 and DHF-ERE-221005).
N.S. acknowledges support from the Knut and Alice Wallenberg foundation through the “Gravity Meets Light” project.
T.T. acknowledges support from the NSF grant AST-2205314 and the NASA ADAP award 80NSSC23K1130.

\section*{Data Availability}

The model is available for public use within {\sc{Redback}}~\citep{Sarin24_redback}.



\bibliographystyle{mnras}
\bibliography{main} 

\begin{thebibliography}{}
\makeatletter
\relax
\def\mn@urlcharsother{\let\do\@makeother \do\$\do\&\do\#\do\^\do\_\do\%\do\~}
\def\mn@doi{\begingroup\mn@urlcharsother \@ifnextchar [ {\mn@doi@}
  {\mn@doi@[]}}
\def\mn@doi@[#1]#2{\def\@tempa{#1}\ifx\@tempa\@empty \href
  {http://dx.doi.org/#2} {doi:#2}\else \href {http://dx.doi.org/#2} {#1}\fi
  \endgroup}
\def\mn@eprint#1#2{\mn@eprint@#1:#2::\@nil}
\def\mn@eprint@arXiv#1{\href {http://arxiv.org/abs/#1} {{\tt arXiv:#1}}}
\def\mn@eprint@dblp#1{\href {http://dblp.uni-trier.de/rec/bibtex/#1.xml}
  {dblp:#1}}
\def\mn@eprint@#1:#2:#3:#4\@nil{\def\@tempa {#1}\def\@tempb {#2}\def\@tempc
  {#3}\ifx \@tempc \@empty \let \@tempc \@tempb \let \@tempb \@tempa \fi \ifx
  \@tempb \@empty \def\@tempb {arXiv}\fi \@ifundefined
  {mn@eprint@\@tempb}{\@tempb:\@tempc}{\expandafter \expandafter \csname
  mn@eprint@\@tempb\endcsname \expandafter{\@tempc}}}

\bibitem[\protect\citeauthoryear{{Angus}, {Levan}, {Perley}  \& et al.}{{Angus}
  et~al.}{2016}]{Angus2016}
{Angus} C.~R.,  {Levan} A.~J.,  {Perley} D.~A.,   et al. 2016, \mn@doi [\mnras]
  {10.1093/mnras/stw063}, \href
  {https://ui.adsabs.harvard.edu/abs/2016MNRAS.458...84A} {458, 84}

\bibitem[\protect\citeauthoryear{{Ashton}, {H{\"u}bner}, {Lasky}  \& et
  al.}{{Ashton} et~al.}{2019}]{Ashton2019}
{Ashton} G.,  {H{\"u}bner} M.,  {Lasky} P.~D.,   et al. 2019, \mn@doi [\apjs]
  {10.3847/1538-4365/ab06fc}, \href
  {https://ui.adsabs.harvard.edu/abs/2019ApJS..241...27A} {241, 27}

\bibitem[\protect\citeauthoryear{{Atoyan}}{{Atoyan}}{1999}]{Atoyan1999}
{Atoyan} A.~M.,  1999, \mn@doi [\aap] {10.48550/arXiv.astro-ph/9905204}, \href
  {https://ui.adsabs.harvard.edu/abs/1999A&A...346L..49A} {346, L49}

\bibitem[\protect\citeauthoryear{{Barr{\`e}re}, {Guilet}, {Reboul-Salze}  \& et
  al.}{{Barr{\`e}re} et~al.}{2022}]{Barrere2022}
{Barr{\`e}re} P.,  {Guilet} J.,  {Reboul-Salze} A.,   et al. 2022, \mn@doi
  [\aap] {10.1051/0004-6361/202244172}, \href
  {https://ui.adsabs.harvard.edu/abs/2022A&A...668A..79B} {668, A79}

\bibitem[\protect\citeauthoryear{{Barr{\`e}re}, {Guilet}, {Raynaud}  \& et
  al.}{{Barr{\`e}re} et~al.}{2023}]{Barrere2023}
{Barr{\`e}re} P.,  {Guilet} J.,  {Raynaud} R.,   et al. 2023, \mn@doi [\mnras]
  {10.1093/mnrasl/slad120}, \href
  {https://ui.adsabs.harvard.edu/abs/2023MNRAS.526L..88B} {526, L88}

\bibitem[\protect\citeauthoryear{{Bietenholz}, {Kronberg}, {Hogg}  \& et
  al.}{{Bietenholz} et~al.}{1991}]{Bietenholz1991}
{Bietenholz} M.~F.,  {Kronberg} P.~P.,  {Hogg} D.~E.,   et al. 1991, \mn@doi
  [\apjl] {10.1086/186051}, \href
  {https://ui.adsabs.harvard.edu/abs/1991ApJ...373L..59B} {373, L59}

\bibitem[\protect\citeauthoryear{{Blanchard}, {Berger}, {Nicholl}  \& et
  al.}{{Blanchard} et~al.}{2020}]{Blanchard2020}
{Blanchard} P.~K.,  {Berger} E.,  {Nicholl} M.,   et al. 2020, \mn@doi [\apj]
  {10.3847/1538-4357/ab9638}, \href
  {https://ui.adsabs.harvard.edu/abs/2020ApJ...897..114B} {897, 114}

\bibitem[\protect\citeauthoryear{{Blondin} \& {Chevalier}}{{Blondin} \&
  {Chevalier}}{2017}]{Blondin2017}
{Blondin} J.~M.,  {Chevalier} R.~A.,  2017, \mn@doi [\apj]
  {10.3847/1538-4357/aa8267}, \href
  {https://ui.adsabs.harvard.edu/abs/2017ApJ...845..139B} {845, 139}

\bibitem[\protect\citeauthoryear{{Borghese}}{{Borghese}}{2023}]{borghese_2020}
{Borghese} A.,  2023, \mn@doi [IAUS] {10.1017/S1743921322000357}, \href
  {https://ui.adsabs.harvard.edu/abs/2023IAUS..363...51B} {363, 51}

\bibitem[\protect\citeauthoryear{{Bucciantini}, {Amato}, {Bandiera}  \& et
  al.}{{Bucciantini} et~al.}{2004}]{Bucciantini2004}
{Bucciantini} N.,  {Amato} E.,  {Bandiera} R.,   et al. 2004, \mn@doi [\aap]
  {10.1051/0004-6361:20040360}, \href
  {https://ui.adsabs.harvard.edu/abs/2004A&A...423..253B} {423, 253}

\bibitem[\protect\citeauthoryear{{Buchner}, {Georgakakis}, {Nandra}  \& et
  al.}{{Buchner} et~al.}{2014}]{Buchner2014}
{Buchner} J.,  {Georgakakis} A.,  {Nandra} K.,   et al. 2014, \mn@doi [\aap]
  {10.1051/0004-6361/201322971}, \href
  {https://ui.adsabs.harvard.edu/abs/2014A&A...564A.125B} {564, A125}

\bibitem[\protect\citeauthoryear{{B{\"u}hler} \& {Blandford}}{{B{\"u}hler} \&
  {Blandford}}{2014}]{Buhler2014}
{B{\"u}hler} R.,  {Blandford} R.,  2014, \mn@doi [Reports on Progress in
  Physics] {10.1088/0034-4885/77/6/066901}, \href
  {https://ui.adsabs.harvard.edu/abs/2014RPPh...77f6901B} {77, 066901}

\bibitem[\protect\citeauthoryear{{Callis}, {Fraser}, {Pastorello}  \& et
  al.}{{Callis} et~al.}{2021}]{Callis2021}
{Callis} E.,  {Fraser} M.,  {Pastorello} A.,   et al. 2021, \mn@doi [arXiv
  e-prints] {10.48550/arXiv.2109.12943}, \href
  {https://ui.adsabs.harvard.edu/abs/2021arXiv210912943C} {p. arXiv:2109.12943}

\bibitem[\protect\citeauthoryear{{Camus}, {Komissarov}, {Bucciantini}  \& et
  al.}{{Camus} et~al.}{2009}]{Camus2009}
{Camus} N.~F.,  {Komissarov} S.~S.,  {Bucciantini} N.,   et al. 2009, \mn@doi
  [\mnras] {10.1111/j.1365-2966.2009.15550.x}, \href
  {https://ui.adsabs.harvard.edu/abs/2009MNRAS.400.1241C} {400, 1241}

\bibitem[\protect\citeauthoryear{{Chen}, {Brennan}, {Wesson}  \& et al.}{{Chen}
  et~al.}{2021}]{Chen2021}
{Chen} T.~W.,  {Brennan} S.~J.,  {Wesson} R.,   et al. 2021, arXiv e-prints,
  \href {https://ui.adsabs.harvard.edu/abs/2021arXiv210907942C} {p.
  arXiv:2109.07942}

\bibitem[\protect\citeauthoryear{{Chen}, {Yan}, {Kangas}  \& et al.}{{Chen}
  et~al.}{2023a}]{Chen2023a}
{Chen} Z.~H.,  {Yan} L.,  {Kangas} T.,   et al. 2023a, \mn@doi [\apj]
  {10.3847/1538-4357/aca161}, \href
  {https://ui.adsabs.harvard.edu/abs/2023ApJ...943...41C} {943, 41}

\bibitem[\protect\citeauthoryear{{Chen}, {Yan}, {Kangas}  \& et al.}{{Chen}
  et~al.}{2023b}]{Chen2023b}
{Chen} Z.~H.,  {Yan} L.,  {Kangas} T.,   et al. 2023b, \mn@doi [\apj]
  {10.3847/1538-4357/aca162}, \href
  {https://ui.adsabs.harvard.edu/abs/2023ApJ...943...42C} {943, 42}

\bibitem[\protect\citeauthoryear{{Chevalier} \& {Fransson}}{{Chevalier} \&
  {Fransson}}{1992}]{Chevalier1992}
{Chevalier} R.~A.,  {Fransson} C.,  1992, \mn@doi [\apj] {10.1086/171674},
  \href {https://ui.adsabs.harvard.edu/abs/1992ApJ...395..540C} {395, 540}

\bibitem[\protect\citeauthoryear{{Clark} \& {Stephenson}}{{Clark} \&
  {Stephenson}}{1977}]{Clark1977}
{Clark} D.~H.,  {Stephenson} F.~R.,  1977, {The historical supernovae}

\bibitem[\protect\citeauthoryear{{Clark}, {Murdin}, {Wood}  \& et al.}{{Clark}
  et~al.}{1983}]{Clark1983}
{Clark} D.~H.,  {Murdin} P.,  {Wood} R.,   et al. 1983, \mn@doi [\mnras]
  {10.1093/mnras/204.2.415}, \href
  {https://ui.adsabs.harvard.edu/abs/1983MNRAS.204..415C} {204, 415}

\bibitem[\protect\citeauthoryear{{Collins}, {Claspy}  \& {Martin}}{{Collins}
  et~al.}{1999}]{Collins1999}
{Collins} George~W. I.,  {Claspy} W.~P.,   {Martin} J.~C.,  1999, \mn@doi
  [\pasp] {10.1086/316401}, \href
  {https://ui.adsabs.harvard.edu/abs/1999PASP..111..871C} {111, 871}

\bibitem[\protect\citeauthoryear{{Crotts}, {Kunkel}  \& {McCarthy}}{{Crotts}
  et~al.}{1989}]{Crotts1989}
{Crotts} A. P.~S.,  {Kunkel} W.~E.,   {McCarthy} P.~J.,  1989, \mn@doi [\apjl]
  {10.1086/185608}, \href
  {https://ui.adsabs.harvard.edu/abs/1989ApJ...347L..61C} {347, L61}

\bibitem[\protect\citeauthoryear{{Davidson} \& {Fesen}}{{Davidson} \&
  {Fesen}}{1985}]{Davidson1985}
{Davidson} K.,  {Fesen} R.~A.,  1985, \mn@doi [\araa]
  {10.1146/annurev.aa.23.090185.001003}, \href
  {https://ui.adsabs.harvard.edu/abs/1985ARA&A..23..119D} {23, 119}

\bibitem[\protect\citeauthoryear{{Dessart}}{{Dessart}}{2019}]{Dessart2019}
{Dessart} L.,  2019, \mn@doi [\aap] {10.1051/0004-6361/201834535}, \href
  {https://ui.adsabs.harvard.edu/abs/2019A&A...621A.141D} {621, A141}

\bibitem[\protect\citeauthoryear{{Dessart}, {Hillier}, {Li}  \& et
  al.}{{Dessart} et~al.}{2012}]{Dessart2012}
{Dessart} L.,  {Hillier} D.~J.,  {Li} C.,   et al. 2012, \mn@doi [\mnras]
  {10.1111/j.1365-2966.2012.21374.x}, \href
  {https://ui.adsabs.harvard.edu/abs/2012MNRAS.424.2139D} {424, 2139}

\bibitem[\protect\citeauthoryear{{Dexter} \& {Kasen}}{{Dexter} \&
  {Kasen}}{2013}]{Dexter2013}
{Dexter} J.,  {Kasen} D.,  2013, \mn@doi [\apj] {10.1088/0004-637X/772/1/30},
  \href {https://ui.adsabs.harvard.edu/abs/2013ApJ...772...30D} {772, 30}

\bibitem[\protect\citeauthoryear{{Eftekhari}, {Berger}, {Margalit}  \& et
  al.}{{Eftekhari} et~al.}{2019}]{Eftekhari2019}
{Eftekhari} T.,  {Berger} E.,  {Margalit} B.,   et al. 2019, \mn@doi [\apjl]
  {10.3847/2041-8213/ab18a5}, \href
  {https://ui.adsabs.harvard.edu/abs/2019ApJ...876L..10E} {876, L10}

\bibitem[\protect\citeauthoryear{{Eftekhari}, {Margalit}, {Omand}  \& et
  al.}{{Eftekhari} et~al.}{2021}]{Eftekhari2021}
{Eftekhari} T.,  {Margalit} B.,  {Omand} C.~M.~B.,   et al. 2021, \mn@doi
  [\apj] {10.3847/1538-4357/abe9b8}, \href
  {https://ui.adsabs.harvard.edu/abs/2021ApJ...912...21E} {912, 21}

\bibitem[\protect\citeauthoryear{{Fesen}, {Shull}  \& {Hurford}}{{Fesen}
  et~al.}{1997}]{Fesen1997}
{Fesen} R.~A.,  {Shull} J.~M.,   {Hurford} A.~P.,  1997, \mn@doi [\aj]
  {10.1086/118258}, \href
  {https://ui.adsabs.harvard.edu/abs/1997AJ....113..354F} {113, 354}

\bibitem[\protect\citeauthoryear{{Frail}, {Kassim}, {Cornwell}  \& et
  al.}{{Frail} et~al.}{1995}]{Frail1995}
{Frail} D.~A.,  {Kassim} N.~E.,  {Cornwell} T.~J.,   et al. 1995, \mn@doi
  [\apjl] {10.1086/309794}, \href
  {https://ui.adsabs.harvard.edu/abs/1995ApJ...454L.129F} {454, L129}

\bibitem[\protect\citeauthoryear{{Fuller}, {Piro}  \& {Jermyn}}{{Fuller}
  et~al.}{2019}]{Fuller2019}
{Fuller} J.,  {Piro} A.~L.,   {Jermyn} A.~S.,  2019, \mn@doi [\mnras]
  {10.1093/mnras/stz514}, \href
  {https://ui.adsabs.harvard.edu/abs/2019MNRAS.485.3661F} {485, 3661}

\bibitem[\protect\citeauthoryear{{Gomez}, {Berger}, {Nicholl}  \& et
  al.}{{Gomez} et~al.}{2022}]{Gomez2022}
{Gomez} S.,  {Berger} E.,  {Nicholl} M.,   et al. 2022, \mn@doi [\apj]
  {10.3847/1538-4357/ac9842}, \href
  {https://ui.adsabs.harvard.edu/abs/2022ApJ...941..107G} {941, 107}

\bibitem[\protect\citeauthoryear{{Gomez}, {Nicholl}, {Berger}  \& et
  al.}{{Gomez} et~al.}{2024}]{Gomez2024}
{Gomez} S.,  {Nicholl} M.,  {Berger} E.,   et al. 2024, \mn@doi [\mnras]
  {10.1093/mnras/stae2270}, \href
  {https://ui.adsabs.harvard.edu/abs/2024MNRAS.535..471G} {535, 471}

\bibitem[\protect\citeauthoryear{{Guilet}, {Reboul-Salze}, {Raynaud}  \& et
  al.}{{Guilet} et~al.}{2022}]{Guilet2022}
{Guilet} J.,  {Reboul-Salze} A.,  {Raynaud} R.,   et al. 2022, \mn@doi [\mnras]
  {10.1093/mnras/stac2499}, \href
  {https://ui.adsabs.harvard.edu/abs/2022MNRAS.516.4346G} {516, 4346}

\bibitem[\protect\citeauthoryear{{Hachinger}, {Mazzali}, {Taubenberger}  \& et
  al.}{{Hachinger} et~al.}{2012}]{Hachinger2012}
{Hachinger} S.,  {Mazzali} P.~A.,  {Taubenberger} S.,   et al. 2012, \mn@doi
  [\mnras] {10.1111/j.1365-2966.2012.20464.x}, \href
  {https://ui.adsabs.harvard.edu/abs/2012MNRAS.422...70H} {422, 70}

\bibitem[\protect\citeauthoryear{{Hester}}{{Hester}}{2008}]{Hester2008}
{Hester} J.~J.,  2008, \mn@doi [\araa]
  {10.1146/annurev.astro.45.051806.110608}, \href
  {https://ui.adsabs.harvard.edu/abs/2008ARA&A..46..127H} {46, 127}

\bibitem[\protect\citeauthoryear{{Hiramatsu}, {Howell}, {Van Dyk}  \& et
  al.}{{Hiramatsu} et~al.}{2021}]{Hiramatsu2021}
{Hiramatsu} D.,  {Howell} D.~A.,  {Van Dyk} S.~D.,   et al. 2021, \mn@doi
  [Nature Astronomy] {10.1038/s41550-021-01384-2}, \href
  {https://ui.adsabs.harvard.edu/abs/2021NatAs...5..903H} {5, 903}

\bibitem[\protect\citeauthoryear{{Igoshev}, {Popov}  \& {Hollerbach}}{{Igoshev}
  et~al.}{2021}]{Igoshev+21}
{Igoshev} A.~P.,  {Popov} S.~B.,   {Hollerbach} R.,  2021, \mn@doi [Univ]
  {10.3390/universe7090351}, \href
  {https://ui.adsabs.harvard.edu/abs/2021Univ....7..351I} {7, 351}

\bibitem[\protect\citeauthoryear{{Inserra}, {Bulla}, {Sim}  \& et
  al.}{{Inserra} et~al.}{2016}]{Inserra2016}
{Inserra} C.,  {Bulla} M.,  {Sim} S.~A.,   et al. 2016, \mn@doi [\apj]
  {10.3847/0004-637X/831/1/79}, \href
  {https://ui.adsabs.harvard.edu/abs/2016ApJ...831...79I} {831, 79}

\bibitem[\protect\citeauthoryear{{Inserra}, {Smartt}, {Gall}  \& et
  al.}{{Inserra} et~al.}{2018}]{Inserra2018}
{Inserra} C.,  {Smartt} S.~J.,  {Gall} E.~E.~E.,   et al. 2018, \mn@doi
  [\mnras] {10.1093/mnras/stx3179}, \href
  {https://ui.adsabs.harvard.edu/abs/2018MNRAS.475.1046I} {475, 1046}

\bibitem[\protect\citeauthoryear{{Jerkstrand}, {Smartt}, {Inserra}  \& et
  al.}{{Jerkstrand} et~al.}{2017}]{Jerkstrand2017}
{Jerkstrand} A.,  {Smartt} S.~J.,  {Inserra} C.,   et al. 2017, \mn@doi [\apj]
  {10.3847/1538-4357/835/1/13}, \href
  {https://ui.adsabs.harvard.edu/abs/2017ApJ...835...13J} {835, 13}

\bibitem[\protect\citeauthoryear{{Jun}}{{Jun}}{1998}]{Jun1998}
{Jun} B.-I.,  1998, \mn@doi [\apj] {10.1086/305627}, \href
  {https://ui.adsabs.harvard.edu/abs/1998ApJ...499..282J} {499, 282}

\bibitem[\protect\citeauthoryear{{Kangas}, {Yan}, {Schulze}  \& et
  al.}{{Kangas} et~al.}{2022}]{Kangas2022}
{Kangas} T.,  {Yan} L.,  {Schulze} S.,   et al. 2022, \mn@doi [\mnras]
  {10.1093/mnras/stac2218}, \href
  {https://ui.adsabs.harvard.edu/abs/2022MNRAS.516.1193K} {516, 1193}

\bibitem[\protect\citeauthoryear{{Kitaura}, {Janka}  \&
  {Hillebrandt}}{{Kitaura} et~al.}{2006}]{Kitaura2006}
{Kitaura} F.~S.,  {Janka} H.~T.,   {Hillebrandt} W.,  2006, \mn@doi [\aap]
  {10.1051/0004-6361:20054703}, \href
  {https://ui.adsabs.harvard.edu/abs/2006A&A...450..345K} {450, 345}

\bibitem[\protect\citeauthoryear{{Kou} \& {Tong}}{{Kou} \&
  {Tong}}{2015}]{Kou2015}
{Kou} F.~F.,  {Tong} H.,  2015, \mn@doi [\mnras] {10.1093/mnras/stv734}, \href
  {https://ui.adsabs.harvard.edu/abs/2015MNRAS.450.1990K} {450, 1990}

\bibitem[\protect\citeauthoryear{{Krause}, {Birkmann}, {Usuda}  \& et
  al.}{{Krause} et~al.}{2008a}]{Krause2008b}
{Krause} O.,  {Birkmann} S.~M.,  {Usuda} T.,   et al. 2008a, \mn@doi [Science]
  {10.1126/science.1155788}, \href
  {https://ui.adsabs.harvard.edu/abs/2008Sci...320.1195K} {320, 1195}

\bibitem[\protect\citeauthoryear{{Krause}, {Tanaka}, {Usuda}  \& et
  al.}{{Krause} et~al.}{2008b}]{Krause2008}
{Krause} O.,  {Tanaka} M.,  {Usuda} T.,   et al. 2008b, \mn@doi [\nat]
  {10.1038/nature07608}, \href
  {https://ui.adsabs.harvard.edu/abs/2008Natur.456..617K} {456, 617}

\bibitem[\protect\citeauthoryear{{Lasky}, {Leris}, {Rowlinson}  \& et
  al.}{{Lasky} et~al.}{2017}]{Lasky2017}
{Lasky} P.~D.,  {Leris} C.,  {Rowlinson} A.,   et al. 2017, \mn@doi [\apjl]
  {10.3847/2041-8213/aa79a7}, \href
  {https://ui.adsabs.harvard.edu/abs/2017ApJ...843L...1L} {843, L1}

\bibitem[\protect\citeauthoryear{{Law}, {Omand}, {Kashiyama}  \& et al.}{{Law}
  et~al.}{2019}]{Law2019}
{Law} C.~J.,  {Omand} C.~M.~B.,  {Kashiyama} K.,   et al. 2019, \mn@doi [\apj]
  {10.3847/1538-4357/ab4adb}, \href
  {https://ui.adsabs.harvard.edu/abs/2019ApJ...886...24L} {886, 24}

\bibitem[\protect\citeauthoryear{{Leloudas}, {Schulze}, {Kr{\"u}hler}  \& et
  al.}{{Leloudas} et~al.}{2015}]{Leloudas2015b}
{Leloudas} G.,  {Schulze} S.,  {Kr{\"u}hler} T.,   et al. 2015, \mn@doi
  [\mnras] {10.1093/mnras/stv320}, \href
  {https://ui.adsabs.harvard.edu/abs/2015MNRAS.449..917L} {449, 917}

\bibitem[\protect\citeauthoryear{{Li}, {Yu}  \& {Huang}}{{Li}
  et~al.}{2015}]{Li2015}
{Li} S.-Z.,  {Yu} Y.-W.,   {Huang} Y.,  2015, \mn@doi [Research in Astronomy
  and Astrophysics] {10.1088/1674-4527/15/11/005}, \href
  {https://ui.adsabs.harvard.edu/abs/2015RAA....15.1823L} {15, 1823}

\bibitem[\protect\citeauthoryear{{Lin}, {van Kerkwijk}, {Kirsten}  \& et
  al.}{{Lin} et~al.}{2023}]{Lin2023}
{Lin} R.,  {van Kerkwijk} M.~H.,  {Kirsten} F.,   et al. 2023, \mn@doi [\apj]
  {10.3847/1538-4357/acdc98}, \href
  {https://ui.adsabs.harvard.edu/abs/2023ApJ...952..161L} {952, 161}

\bibitem[\protect\citeauthoryear{{Lunnan}, {Chornock}, {Berger}  \& et
  al.}{{Lunnan} et~al.}{2014}]{Lunnan2014}
{Lunnan} R.,  {Chornock} R.,  {Berger} E.,   et al. 2014, \mn@doi [\apj]
  {10.1088/0004-637X/787/2/138}, \href
  {https://ui.adsabs.harvard.edu/abs/2014ApJ...787..138L} {787, 138}

\bibitem[\protect\citeauthoryear{{Lyne}, {Pritchard}  \& {Graham Smith}}{{Lyne}
  et~al.}{1993}]{Lyne1993}
{Lyne} A.~G.,  {Pritchard} R.~S.,   {Graham Smith} F.,  1993, \mn@doi [\mnras]
  {10.1093/mnras/265.4.1003}, \href
  {https://ui.adsabs.harvard.edu/abs/1993MNRAS.265.1003L} {265, 1003}

\bibitem[\protect\citeauthoryear{{Lyne}, {Jordan}, {Graham-Smith}  \& et
  al.}{{Lyne} et~al.}{2015}]{Lyne2015}
{Lyne} A.~G.,  {Jordan} C.~A.,  {Graham-Smith} F.,   et al. 2015, \mn@doi
  [\mnras] {10.1093/mnras/stu2118}, \href
  {https://ui.adsabs.harvard.edu/abs/2015MNRAS.446..857L} {446, 857}

\bibitem[\protect\citeauthoryear{{MacAlpine}, {McGaugh}, {Mazzarella}  \& et
  al.}{{MacAlpine} et~al.}{1989}]{MacAlpine1989}
{MacAlpine} G.~M.,  {McGaugh} S.~S.,  {Mazzarella} J.~M.,   et al. 1989,
  \mn@doi [\apj] {10.1086/167598}, \href
  {https://ui.adsabs.harvard.edu/abs/1989ApJ...342..364M} {342, 364}

\bibitem[\protect\citeauthoryear{{Manchester}, {Hobbs}, {Teoh}  \&
  {Hobbs}}{{Manchester} et~al.}{2005}]{Manchester2005}
{Manchester} R.~N.,  {Hobbs} G.~B.,  {Teoh} A.,   {Hobbs} M.,  2005, \mn@doi
  [\aj] {10.1086/428488}, \href
  {https://ui.adsabs.harvard.edu/abs/2005AJ....129.1993M} {129, 1993}

\bibitem[\protect\citeauthoryear{{Margutti}, {Bright}, {Matthews}  \& et
  al.}{{Margutti} et~al.}{2023}]{Margutti2023}
{Margutti} R.,  {Bright} J.~S.,  {Matthews} D.~J.,   et al. 2023, \mn@doi
  [\apjl] {10.3847/2041-8213/acf1fd}, \href
  {https://ui.adsabs.harvard.edu/abs/2023ApJ...954L..45M} {954, L45}

\bibitem[\protect\citeauthoryear{{Mauche} \& {Gorenstein}}{{Mauche} \&
  {Gorenstein}}{1989}]{Mauche1989}
{Mauche} C.~W.,  {Gorenstein} P.,  1989, \mn@doi [\apj] {10.1086/167055}, \href
  {https://ui.adsabs.harvard.edu/abs/1989ApJ...336..843M} {336, 843}

\bibitem[\protect\citeauthoryear{{Melatos}}{{Melatos}}{1999}]{Melatos1999}
{Melatos} A.,  1999, \mn@doi [\apjl] {10.1086/312104}, \href
  {https://ui.adsabs.harvard.edu/abs/1999ApJ...519L..77M} {519, L77}

\bibitem[\protect\citeauthoryear{{Metzger}}{{Metzger}}{2019}]{Metzger2019}
{Metzger} B.~D.,  2019, \mn@doi [Living Reviews in Relativity]
  {10.1007/s41114-019-0024-0}, \href
  {https://ui.adsabs.harvard.edu/abs/2019LRR....23....1M} {23, 1}

\bibitem[\protect\citeauthoryear{{Miller}}{{Miller}}{1973}]{Miller1973}
{Miller} J.~S.,  1973, \mn@doi [\apjl] {10.1086/181158}, \href
  {https://ui.adsabs.harvard.edu/abs/1973ApJ...180L..83M} {180, L83}

\bibitem[\protect\citeauthoryear{{Miyaji}, {Nomoto}, {Yokoi}  \& et
  al.}{{Miyaji} et~al.}{1980}]{Miyaji1980}
{Miyaji} S.,  {Nomoto} K.,  {Yokoi} K.,   et al. 1980, \pasj, \href
  {https://ui.adsabs.harvard.edu/abs/1980PASJ...32..303M} {32, 303}

\bibitem[\protect\citeauthoryear{{Mondal}, {Bera}, {Chandra}  \& et
  al.}{{Mondal} et~al.}{2020}]{Mondal2020}
{Mondal} S.,  {Bera} A.,  {Chandra} P.,   et al. 2020, \mn@doi [\mnras]
  {10.1093/mnras/staa2637}, \href
  {https://ui.adsabs.harvard.edu/abs/2020MNRAS.498.3863M} {498, 3863}

\bibitem[\protect\citeauthoryear{{Murase}, {Kashiyama}, {Kiuchi}  \& et
  al.}{{Murase} et~al.}{2015}]{Murase2015}
{Murase} K.,  {Kashiyama} K.,  {Kiuchi} K.,   et al. 2015, \mn@doi [\apj]
  {10.1088/0004-637X/805/1/82}, \href
  {https://ui.adsabs.harvard.edu/abs/2015ApJ...805...82M} {805, 82}

\bibitem[\protect\citeauthoryear{{Nicholl}, {Guillochon}  \&
  {Berger}}{{Nicholl} et~al.}{2017}]{Nicholl2017}
{Nicholl} M.,  {Guillochon} J.,   {Berger} E.,  2017, \mn@doi [\apj]
  {10.3847/1538-4357/aa9334}, \href
  {https://ui.adsabs.harvard.edu/abs/2017ApJ...850...55N} {850, 55}

\bibitem[\protect\citeauthoryear{{Nomoto}}{{Nomoto}}{1987}]{Nomoto1987}
{Nomoto} K.,  1987, \mn@doi [\apj] {10.1086/165716}, \href
  {https://ui.adsabs.harvard.edu/abs/1987ApJ...322..206N} {322, 206}

\bibitem[\protect\citeauthoryear{{Nomoto}, {Sparks}, {Fesen}  \& et
  al.}{{Nomoto} et~al.}{1982}]{Nomoto1982}
{Nomoto} K.,  {Sparks} W.~M.,  {Fesen} R.~A.,   et al. 1982, \mn@doi [\nat]
  {10.1038/299803a0}, \href
  {https://ui.adsabs.harvard.edu/abs/1982Natur.299..803N} {299, 803}

\bibitem[\protect\citeauthoryear{{Omand} \& {Jerkstrand}}{{Omand} \&
  {Jerkstrand}}{2023}]{Omand2023}
{Omand} C.~M.~B.,  {Jerkstrand} A.,  2023, \mn@doi [\aap]
  {10.1051/0004-6361/202245406}, \href
  {https://ui.adsabs.harvard.edu/abs/2023A&A...673A.107O} {673, A107}

\bibitem[\protect\citeauthoryear{{Omand} \& {Sarin}}{{Omand} \&
  {Sarin}}{2024}]{Omand2024}
{Omand} C. M.~B.,  {Sarin} N.,  2024, \mn@doi [\mnras]
  {10.1093/mnras/stad3645}, \href
  {https://ui.adsabs.harvard.edu/abs/2024MNRAS.527.6455O} {527, 6455}

\bibitem[\protect\citeauthoryear{{Omand}, {Kashiyama}  \& {Murase}}{{Omand}
  et~al.}{2018}]{Omand2018}
{Omand} C. M.~B.,  {Kashiyama} K.,   {Murase} K.,  2018, \mn@doi [\mnras]
  {10.1093/mnras/stx2743}, \href
  {https://ui.adsabs.harvard.edu/abs/2018MNRAS.474..573O} {474, 573}

\bibitem[\protect\citeauthoryear{{Omand}, {Kashiyama}  \& {Murase}}{{Omand}
  et~al.}{2019}]{Omand2019}
{Omand} C. M.~B.,  {Kashiyama} K.,   {Murase} K.,  2019, \mn@doi [\mnras]
  {10.1093/mnras/stz371}, \href
  {https://ui.adsabs.harvard.edu/abs/2019MNRAS.484.5468O} {484, 5468}

\bibitem[\protect\citeauthoryear{{{\O}rum}, {Ivens}, {Strandberg}  \& et
  al.}{{{\O}rum} et~al.}{2020}]{Orum2020}
{{\O}rum} S.~V.,  {Ivens} D.~L.,  {Strandberg} P.,   et al. 2020, \mn@doi
  [\aap] {10.1051/0004-6361/202038176}, \href
  {https://ui.adsabs.harvard.edu/abs/2020A&A...643A..47O} {643, A47}

\bibitem[\protect\citeauthoryear{{Owen} \& {Barlow}}{{Owen} \&
  {Barlow}}{2015}]{Owen2015}
{Owen} P.~J.,  {Barlow} M.~J.,  2015, \mn@doi [\apj]
  {10.1088/0004-637X/801/2/141}, \href
  {https://ui.adsabs.harvard.edu/abs/2015ApJ...801..141O} {801, 141}

\bibitem[\protect\citeauthoryear{{Pacini}}{{Pacini}}{1967}]{Pacini1967}
{Pacini} F.,  1967, \mn@doi [\nat] {10.1038/216567a0}, \href
  {https://ui.adsabs.harvard.edu/abs/1967Natur.216..567P} {216, 567}

\bibitem[\protect\citeauthoryear{{Pessi}, {Anderson}, {Folatelli}  \& et
  al.}{{Pessi} et~al.}{2023}]{Pessi2023}
{Pessi} P.~J.,  {Anderson} J.~P.,  {Folatelli} G.,   et al. 2023, \mn@doi
  [\mnras] {10.1093/mnras/stad1822}, \href
  {https://ui.adsabs.harvard.edu/abs/2023MNRAS.523.5315P} {523, 5315}

\bibitem[\protect\citeauthoryear{{Pitkin}}{{Pitkin}}{2018}]{psrqpy}
{Pitkin} M.,  2018, \mn@doi [{JOSS}] {10.21105/joss.00538}, 3, 538

\bibitem[\protect\citeauthoryear{{Poidevin}, {Omand}, {P{\'e}rez-Fournon}  \&
  et al.}{{Poidevin} et~al.}{2022}]{Poidevin2022}
{Poidevin} F.,  {Omand} C.~M.~B.,  {P{\'e}rez-Fournon} I.,   et al. 2022,
  \mn@doi [\mnras] {10.1093/mnras/stac425}, \href
  {https://ui.adsabs.harvard.edu/abs/2022MNRAS.511.5948P} {511, 5948}

\bibitem[\protect\citeauthoryear{{Poidevin}, {Omand}, {K{\"o}nyves-T{\'o}th}
  \& et al.}{{Poidevin} et~al.}{2023}]{Poidevin2023}
{Poidevin} F.,  {Omand} C.~M.~B.,  {K{\"o}nyves-T{\'o}th} R.,   et al. 2023,
  \mn@doi [\mnras] {10.1093/mnras/stad830}, \href
  {https://ui.adsabs.harvard.edu/abs/2023MNRAS.521.5418P} {521, 5418}

\bibitem[\protect\citeauthoryear{{Pons} \& {Geppert}}{{Pons} \&
  {Geppert}}{2007}]{Pons2007}
{Pons} J.~A.,  {Geppert} U.,  2007, \mn@doi [\aap]
  {10.1051/0004-6361:20077456}, \href
  {https://ui.adsabs.harvard.edu/abs/2007A&A...470..303P} {470, 303}

\bibitem[\protect\citeauthoryear{{Popov}}{{Popov}}{1993}]{Popov1993}
{Popov} D.~V.,  1993, \mn@doi [\apj] {10.1086/173117}, \href
  {https://ui.adsabs.harvard.edu/abs/1993ApJ...414..712P} {414, 712}

\bibitem[\protect\citeauthoryear{{Porth}, {Komissarov}  \& {Keppens}}{{Porth}
  et~al.}{2014}]{Porth2014}
{Porth} O.,  {Komissarov} S.~S.,   {Keppens} R.,  2014, \mn@doi [\mnras]
  {10.1093/mnras/stu1082}, \href
  {https://ui.adsabs.harvard.edu/abs/2014MNRAS.443..547P} {443, 547}

\bibitem[\protect\citeauthoryear{{Predehl} \& {Schmitt}}{{Predehl} \&
  {Schmitt}}{1995}]{Predehl1995}
{Predehl} P.,  {Schmitt} J.~H.~M.~M.,  1995, \aap, \href
  {https://ui.adsabs.harvard.edu/abs/1995A&A...293..889P} {293, 889}

\bibitem[\protect\citeauthoryear{{Pursiainen}, {Leloudas}, {Paraskeva}  \& et
  al.}{{Pursiainen} et~al.}{2022}]{Pursiainen2022}
{Pursiainen} M.,  {Leloudas} G.,  {Paraskeva} E.,   et al. 2022, \mn@doi [\aap]
  {10.1051/0004-6361/202243256}, \href
  {https://ui.adsabs.harvard.edu/abs/2022A&A...666A..30P} {666, A30}

\bibitem[\protect\citeauthoryear{{Pursiainen}, {Leloudas}, {Cikota}  \& et
  al.}{{Pursiainen} et~al.}{2023}]{Pursiainen2023}
{Pursiainen} M.,  {Leloudas} G.,  {Cikota} A.,   et al. 2023, \mn@doi [\aap]
  {10.1051/0004-6361/202345945}, \href
  {https://ui.adsabs.harvard.edu/abs/2023A&A...674A..81P} {674, A81}

\bibitem[\protect\citeauthoryear{{Raynaud}, {Guilet}, {Janka}  \& et
  al.}{{Raynaud} et~al.}{2020}]{Raynaud2020}
{Raynaud} R.,  {Guilet} J.,  {Janka} H.-T.,   et al. 2020, \mn@doi [Science
  Advances] {10.1126/sciadv.aay2732}, \href
  {https://ui.adsabs.harvard.edu/abs/2020SciA....6.2732R} {6, eaay2732}

\bibitem[\protect\citeauthoryear{{Reboul-Salze}, {Guilet}, {Raynaud}  \&
  {Bugli}}{{Reboul-Salze} et~al.}{2021}]{Reboul-Salze2021}
{Reboul-Salze} A.,  {Guilet} J.,  {Raynaud} R.,   {Bugli} M.,  2021, \mn@doi
  [\aap] {10.1051/0004-6361/202038369}, \href
  {https://ui.adsabs.harvard.edu/abs/2021A&A...645A.109R} {645, A109}

\bibitem[\protect\citeauthoryear{{Reboul-Salze}, {Guilet}, {Raynaud}  \& et
  al.}{{Reboul-Salze} et~al.}{2022}]{Reboul-Salze2022}
{Reboul-Salze} A.,  {Guilet} J.,  {Raynaud} R.,   et al. 2022, \mn@doi [\aap]
  {10.1051/0004-6361/202142368}, \href
  {https://ui.adsabs.harvard.edu/abs/2022A&A...667A..94R} {667, A94}

\bibitem[\protect\citeauthoryear{{Rest}, {Suntzeff}, {Olsen}  \& et al.}{{Rest}
  et~al.}{2005}]{Rest2005}
{Rest} A.,  {Suntzeff} N.~B.,  {Olsen} K.,   et al. 2005, \mn@doi [\nat]
  {10.1038/nature04365}, \href
  {https://ui.adsabs.harvard.edu/abs/2005Natur.438.1132R} {438, 1132}

\bibitem[\protect\citeauthoryear{{Rest}, {Welch}, {Suntzeff}  \& et al.}{{Rest}
  et~al.}{2008}]{Rest2008a}
{Rest} A.,  {Welch} D.~L.,  {Suntzeff} N.~B.,   et al. 2008, \mn@doi [\apjl]
  {10.1086/590427}, \href
  {https://ui.adsabs.harvard.edu/abs/2008ApJ...681L..81R} {681, L81}

\bibitem[\protect\citeauthoryear{{Rest}, {Sinnott}, {Welch}  \& et al.}{{Rest}
  et~al.}{2011a}]{Rest2011a}
{Rest} A.,  {Sinnott} B.,  {Welch} D.~L.,   et al. 2011a, \mn@doi [\apj]
  {10.1088/0004-637X/732/1/2}, \href
  {https://ui.adsabs.harvard.edu/abs/2011ApJ...732....2R} {732, 2}

\bibitem[\protect\citeauthoryear{{Rest}, {Foley}, {Sinnott}  \& et al.}{{Rest}
  et~al.}{2011b}]{Rest2011b}
{Rest} A.,  {Foley} R.~J.,  {Sinnott} B.,   et al. 2011b, \mn@doi [\apj]
  {10.1088/0004-637X/732/1/3}, \href
  {https://ui.adsabs.harvard.edu/abs/2011ApJ...732....3R} {732, 3}

\bibitem[\protect\citeauthoryear{{Rodr{\'\i}guez}, {Nakar}  \&
  {Maoz}}{{Rodr{\'\i}guez} et~al.}{2024}]{Rodriguez2024}
{Rodr{\'\i}guez} {\'O}.,  {Nakar} E.,   {Maoz} D.,  2024, \mn@doi [\nat]
  {10.1038/s41586-024-07262-x}, \href
  {https://ui.adsabs.harvard.edu/abs/2024Natur.628..733R} {628, 733}

\bibitem[\protect\citeauthoryear{{Romani}, {Reach}, {Koo}  \& et al.}{{Romani}
  et~al.}{1990}]{Romani1990}
{Romani} R.~W.,  {Reach} W.~T.,  {Koo} B.~C.,   et al. 1990, \mn@doi [\apjl]
  {10.1086/185649}, \href
  {https://ui.adsabs.harvard.edu/abs/1990ApJ...349L..51R} {349, L51}

\bibitem[\protect\citeauthoryear{{Saito}, {Tanaka}, {Moriya}  \& et
  al.}{{Saito} et~al.}{2020}]{Saito2020}
{Saito} S.,  {Tanaka} M.,  {Moriya} T.~J.,   et al. 2020, \mn@doi [\apj]
  {10.3847/1538-4357/ab873b}, \href
  {https://ui.adsabs.harvard.edu/abs/2020ApJ...894..154S} {894, 154}

\bibitem[\protect\citeauthoryear{{Sarin}, {Lasky}, {Sammut}  \& et al.}{{Sarin}
  et~al.}{2018}]{Sarin2018}
{Sarin} N.,  {Lasky} P.~D.,  {Sammut} L.,   et al. 2018, \mn@doi [\prd]
  {10.1103/PhysRevD.98.043011}, \href
  {https://ui.adsabs.harvard.edu/abs/2018PhRvD..98d3011S} {98, 043011}

\bibitem[\protect\citeauthoryear{{Sarin}, {Lasky}  \& {Ashton}}{{Sarin}
  et~al.}{2020}]{Sarin2020b}
{Sarin} N.,  {Lasky} P.~D.,   {Ashton} G.,  2020, \mn@doi [\prd]
  {10.1103/PhysRevD.101.063021}, \href
  {https://ui.adsabs.harvard.edu/abs/2020PhRvD.101f3021S} {101, 063021}

\bibitem[\protect\citeauthoryear{{Sarin}, {Omand}, {Margalit}  \& et
  al.}{{Sarin} et~al.}{2022}]{Sarin2022}
{Sarin} N.,  {Omand} C. M.~B.,  {Margalit} B.,   et al. 2022, \mn@doi [\mnras]
  {10.1093/mnras/stac2609}, \href
  {https://ui.adsabs.harvard.edu/abs/2022MNRAS.516.4949S} {516, 4949}

\bibitem[\protect\citeauthoryear{{Sarin}, {Brandenburg}  \& {Haskell}}{{Sarin}
  et~al.}{2023}]{Sarin2023_cascades}
{Sarin} N.,  {Brandenburg} A.,   {Haskell} B.,  2023, \mn@doi [\apjl]
  {10.3847/2041-8213/ace363}, \href
  {https://ui.adsabs.harvard.edu/abs/2023ApJ...952L..21S} {952, L21}

\bibitem[\protect\citeauthoryear{{Sarin}, {H{\"u}bner}, {Omand}  \& et
  al.}{{Sarin} et~al.}{2024}]{Sarin24_redback}
{Sarin} N.,  {H{\"u}bner} M.,  {Omand} C. M.~B.,   et al. 2024, \mn@doi
  [\mnras] {10.1093/mnras/stae1238}, \href
  {https://ui.adsabs.harvard.edu/abs/2024MNRAS.531.1203S} {531, 1203}

\bibitem[\protect\citeauthoryear{{Schramm}}{{Schramm}}{1977}]{Schramm1977}
{Schramm} D.,  1977, in Supernovae. , \mn@doi{10.1007/978-94-010-1229-4}

\bibitem[\protect\citeauthoryear{{Schulze}, {Kr{\"u}hler}, {Leloudas}  \& et
  al.}{{Schulze} et~al.}{2018}]{Schulze2018}
{Schulze} S.,  {Kr{\"u}hler} T.,  {Leloudas} G.,   et al. 2018, \mn@doi
  [\mnras] {10.1093/mnras/stx2352}, \href
  {https://ui.adsabs.harvard.edu/abs/2018MNRAS.473.1258S} {473, 1258}

\bibitem[\protect\citeauthoryear{{Seward}, {Gorenstein}  \& {Smith}}{{Seward}
  et~al.}{2006}]{Seward2006}
{Seward} F.~D.,  {Gorenstein} P.,   {Smith} R.~K.,  2006, \mn@doi [\apj]
  {10.1086/498105}, \href
  {https://ui.adsabs.harvard.edu/abs/2006ApJ...636..873S} {636, 873}

\bibitem[\protect\citeauthoryear{{Smith}}{{Smith}}{2003}]{Smith2003}
{Smith} N.,  2003, \mn@doi [\mnras] {10.1111/j.1365-2966.2003.07135.x}, \href
  {https://ui.adsabs.harvard.edu/abs/2003MNRAS.346..885S} {346, 885}

\bibitem[\protect\citeauthoryear{{Smith}}{{Smith}}{2013}]{Smith2013}
{Smith} N.,  2013, \mn@doi [\mnras] {10.1093/mnras/stt1004}, \href
  {https://ui.adsabs.harvard.edu/abs/2013MNRAS.434..102S} {434, 102}

\bibitem[\protect\citeauthoryear{{Sollerman}, {Lundqvist}, {Lindler}  \& et
  al.}{{Sollerman} et~al.}{2000}]{Sollerman2000}
{Sollerman} J.,  {Lundqvist} P.,  {Lindler} D.,   et al. 2000, \mn@doi [\apj]
  {10.1086/309062}, \href
  {https://ui.adsabs.harvard.edu/abs/2000ApJ...537..861S} {537, 861}

\bibitem[\protect\citeauthoryear{{Sollerman}, {Kozma}  \&
  {Lundqvist}}{{Sollerman} et~al.}{2001}]{Sollerman2001}
{Sollerman} J.,  {Kozma} C.,   {Lundqvist} P.,  2001, \mn@doi [\aap]
  {10.1051/0004-6361:20000211}, \href
  {https://ui.adsabs.harvard.edu/abs/2001A&A...366..197S} {366, 197}

\bibitem[\protect\citeauthoryear{{Spruit}}{{Spruit}}{2002}]{Spruit2002}
{Spruit} H.~C.,  2002, \mn@doi [\aap] {10.1051/0004-6361:20011465}, \href
  {https://ui.adsabs.harvard.edu/abs/2002A&A...381..923S} {381, 923}

\bibitem[\protect\citeauthoryear{{Staelin} \& {Reifenstein}}{{Staelin} \&
  {Reifenstein}}{1968}]{Staelin1968}
{Staelin} D.~H.,  {Reifenstein} Edward~C. I.,  1968, \mn@doi [Science]
  {10.1126/science.162.3861.1481}, \href
  {https://ui.adsabs.harvard.edu/abs/1968Sci...162.1481S} {162, 1481}

\bibitem[\protect\citeauthoryear{{Sugerman}, {Ercolano}, {Barlow}  \& et
  al.}{{Sugerman} et~al.}{2006}]{Sugerman2006}
{Sugerman} B. E.~K.,  {Ercolano} B.,  {Barlow} M.~J.,   et al. 2006, \mn@doi
  [Science] {10.1126/science.1128131}, \href
  {https://ui.adsabs.harvard.edu/abs/2006Sci...313..196S} {313, 196}

\bibitem[\protect\citeauthoryear{{Sun}, {Xiao}  \& {Li}}{{Sun}
  et~al.}{2022}]{Sun2022}
{Sun} L.,  {Xiao} L.,   {Li} G.,  2022, \mn@doi [\mnras]
  {10.1093/mnras/stac1121}, \href
  {https://ui.adsabs.harvard.edu/abs/2022MNRAS.513.4057S} {513, 4057}

\bibitem[\protect\citeauthoryear{{Suzuki} \& {Maeda}}{{Suzuki} \&
  {Maeda}}{2017}]{Suzuki2017}
{Suzuki} A.,  {Maeda} K.,  2017, \mn@doi [\mnras] {10.1093/mnras/stw3259},
  \href {https://ui.adsabs.harvard.edu/abs/2017MNRAS.466.2633S} {466, 2633}

\bibitem[\protect\citeauthoryear{{Temim} et~al.,}{{Temim}
  et~al.}{2006}]{Temim2006}
{Temim} T.,  et~al., 2006, \mn@doi [\aj] {10.1086/507076}, \href
  {https://ui.adsabs.harvard.edu/abs/2006AJ....132.1610T} {132, 1610}

\bibitem[\protect\citeauthoryear{{Temim}, {Laming}, {Kavanagh}  \& et
  al.}{{Temim} et~al.}{2024}]{Temim2024}
{Temim} T.,  {Laming} J.~M.,  {Kavanagh} P.~J.,   et al. 2024, \mn@doi [\apjl]
  {10.3847/2041-8213/ad50d1}, \href
  {https://ui.adsabs.harvard.edu/abs/2024ApJ...968L..18T} {968, L18}

\bibitem[\protect\citeauthoryear{{Tsuna}, {Matsumoto}, {Wu}  \& et al.}{{Tsuna}
  et~al.}{2024}]{Tsuna2024}
{Tsuna} D.,  {Matsumoto} T.,  {Wu} S.~C.,   et al. 2024, \mn@doi [\apj]
  {10.3847/1538-4357/ad3637}, \href
  {https://ui.adsabs.harvard.edu/abs/2024ApJ...966...30T} {966, 30}

\bibitem[\protect\citeauthoryear{{Vurm} \& {Metzger}}{{Vurm} \&
  {Metzger}}{2021}]{Vurm2021}
{Vurm} I.,  {Metzger} B.~D.,  2021, \mn@doi [\apj] {10.3847/1538-4357/ac0826},
  \href {https://ui.adsabs.harvard.edu/abs/2021ApJ...917...77V} {917, 77}

\bibitem[\protect\citeauthoryear{{Wallace}, {Landecker}  \& {Taylor}}{{Wallace}
  et~al.}{1994}]{Wallace1994}
{Wallace} B.~J.,  {Landecker} T.~L.,   {Taylor} A.~R.,  1994, \aap, \href
  {https://ui.adsabs.harvard.edu/abs/1994A&A...286..565W} {286, 565}

\bibitem[\protect\citeauthoryear{{Wallace}, {Landecker}, {Kalberla}  \& et
  al.}{{Wallace} et~al.}{1999}]{Wallace1999}
{Wallace} B.~J.,  {Landecker} T.~L.,  {Kalberla} P.~M.~W.,   et al. 1999,
  \mn@doi [\apjs] {10.1086/313254}, \href
  {https://ui.adsabs.harvard.edu/abs/1999ApJS..124..181W} {124, 181}

\bibitem[\protect\citeauthoryear{{Wang} et~al.}{{Wang} et~al.}{2015}]{Wang2015}
{Wang} S.~Q.,  et~al., 2015, \mn@doi [\apj] {10.1088/0004-637X/799/1/107},
  \href {https://ui.adsabs.harvard.edu/abs/2015ApJ...799..107W} {799, 107}

\bibitem[\protect\citeauthoryear{{Yang} \& {Chevalier}}{{Yang} \&
  {Chevalier}}{2015}]{Yang2015}
{Yang} H.,  {Chevalier} R.~A.,  2015, \mn@doi [\apj]
  {10.1088/0004-637X/806/2/153}, \href
  {https://ui.adsabs.harvard.edu/abs/2015ApJ...806..153Y} {806, 153}

\bibitem[\protect\citeauthoryear{{Yu}, {Zhang}  \& {Gao}}{{Yu}
  et~al.}{2013}]{Yu2013}
{Yu} Y.-W.,  {Zhang} B.,   {Gao} H.,  2013, \mn@doi [\apjl]
  {10.1088/2041-8205/776/2/L40}, \href
  {https://ui.adsabs.harvard.edu/abs/Yu2013} {776, L40}

\bibitem[\protect\citeauthoryear{{Zhang}, {Wang}, {J{\'o}zsef}  \& et
  al.}{{Zhang} et~al.}{2020}]{Zhang2020}
{Zhang} J.,  {Wang} X.,  {J{\'o}zsef} V.,   et al. 2020, \mn@doi [\mnras]
  {10.1093/mnras/staa2273}, \href
  {https://ui.adsabs.harvard.edu/abs/2020MNRAS.498...84Z} {498, 84}

\makeatother
\end{thebibliography}




\appendix

\section{Parameter Posteriors for SN 1054} \label{sec:appcorner}

\begin{figure*}
\includegraphics[width=0.9\linewidth]{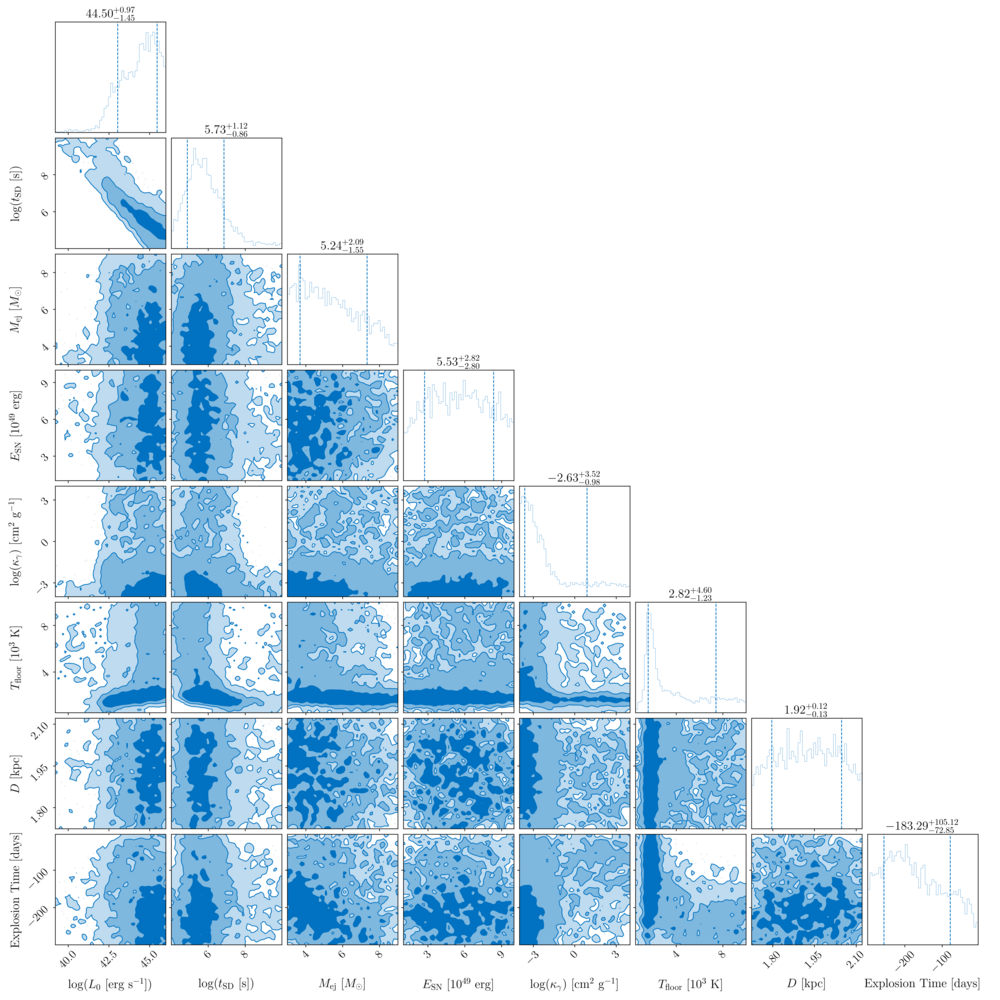}
\caption{Posterior distribution of parameters inferred for SN 1054 for $n = 3$ The explosion time is from when the supernova fades from the daytime sky.}  %
\label{fig:corner}
\end{figure*}

\begin{figure*}
\includegraphics[width=0.9\linewidth]{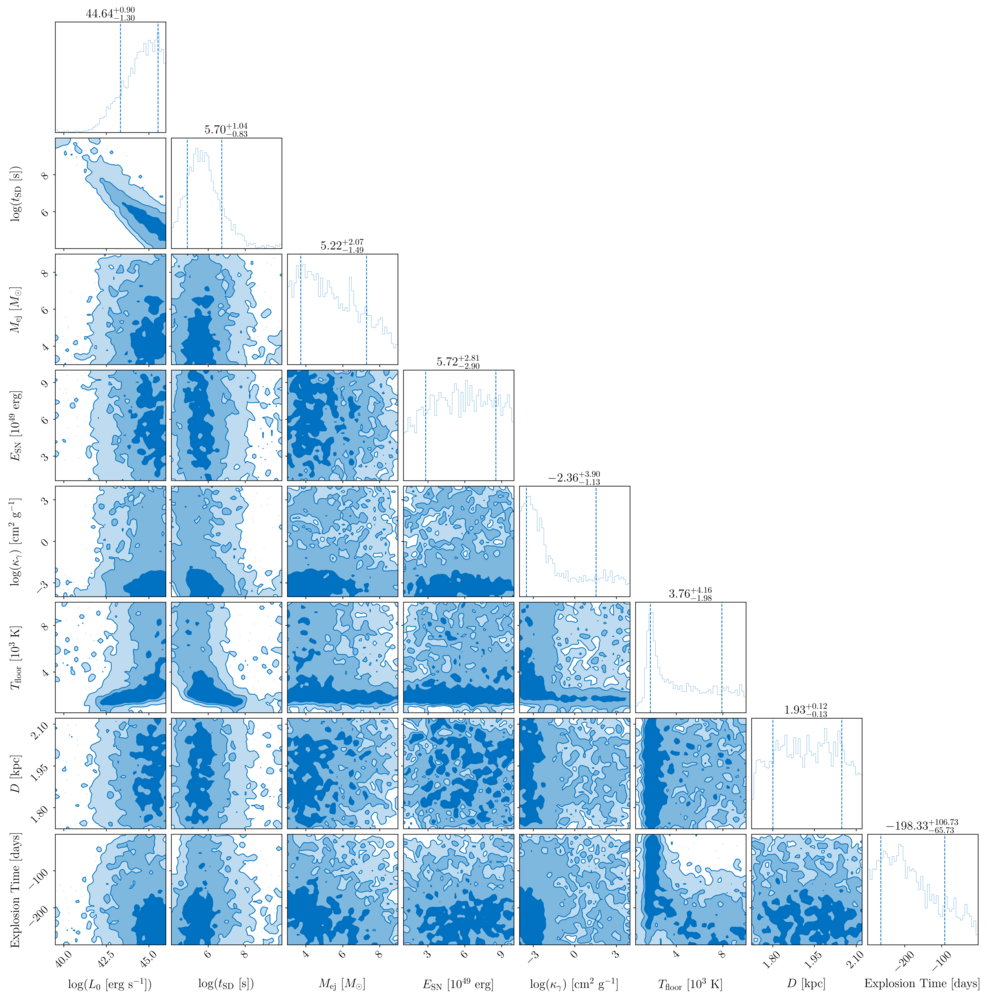}
\caption{Posterior distribution of parameters inferred for SN 1054 for $n = 2.51$.  The explosion time is from when the supernova fades from the daytime sky.}%
\label{fig:cornern251}
\end{figure*}

\begin{figure*}
\includegraphics[width=0.9\linewidth]{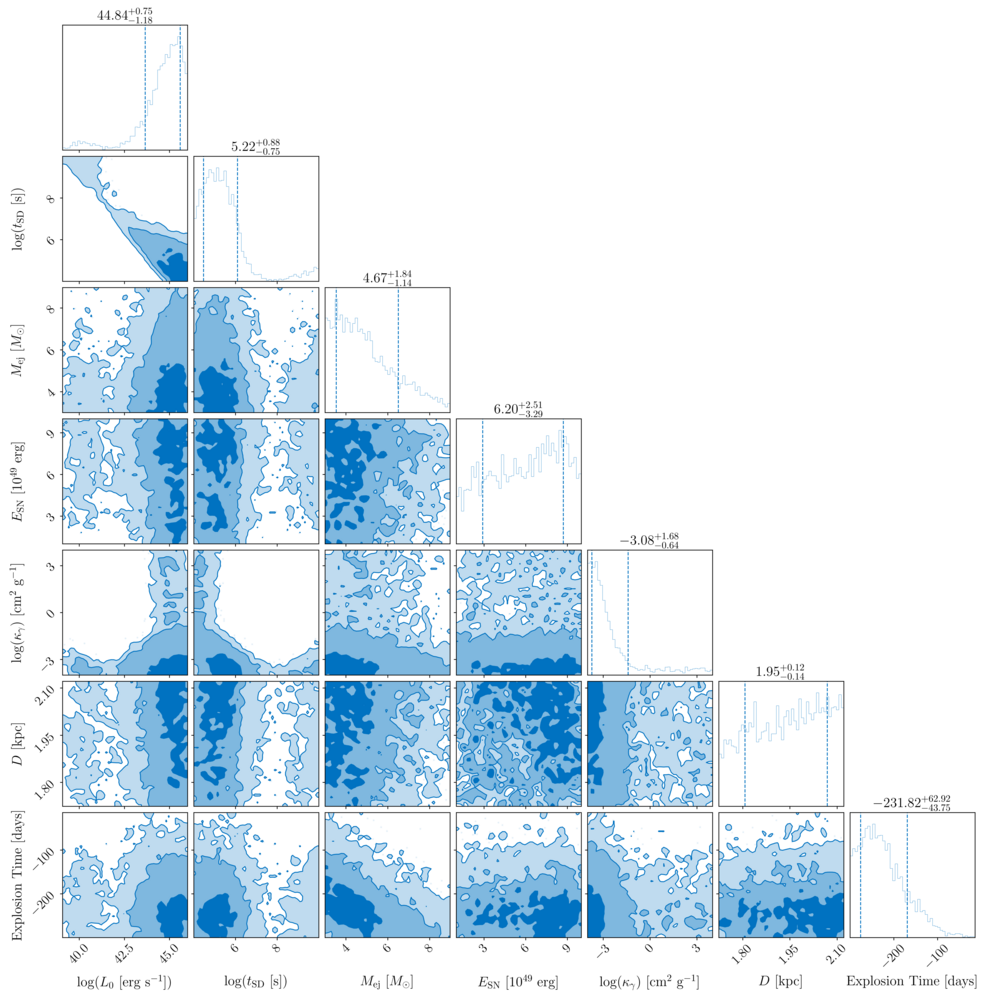}
\caption{Posterior distribution of parameters inferred for SN 1054 for $n = 3$ and $T_{\rm min}$ fixed to 6000 K.  The explosion time is from when the supernova fades from the daytime sky.} %
\label{fig:cornert6000}
\end{figure*}

The full posterior for all inferred parameters when assuming $n = 3$ is shown in Figure \ref{fig:corner}, the posterior when assuming the current Crab value of $n = 2.51$ is shown in \ref{fig:cornern251}, and the posterior when assuming $n = 3$ and $T_{\rm min} = 6000$ K (around the recombination temperature for hydrogen) is shown in Figure \ref{fig:cornert6000}.  The three posteriors show extremely similar behavior, verifying that our results are not strongly affected by these assumptions..  The posteriors for ejecta mass, explosion energy, and distance are almost flat, meaning that little information about these properties can be derived from the light curve.  The posteriors for explosion time and gamma ray opacity all tend towards lower values, but are still broad enough that the value we infer is not well constrained.  The temperature when the photosphere recedes is well constrained to around $\sim$ 3000 -- 4000 K.

\section{Analytical Estimation of Initial Spin Frequency} \label{sec:appspin}

We can take a simple extrapolation from only currently observed values to derive an estimate of the initial spin period.
By taking the equation for spin down,

\begin{equation}
    \dot{\nu} = -k\nu^n ,
    \label{eqn:dotnuode}
\end{equation}
and assuming constant $k$ and braking index $n$, we can get an estimate of the initial spin period of the pulsar.  Solving for $\nu$ and $\dot{\nu}$ gives

\begin{align}
    \nu = & \big[(n-1)(C_1 + kt)\big]^{\frac{1}{1-n}}, \label{eqn:nu} \\
    \dot{\nu} = & -k\big[(n-1)(C_1 + kt)\big]^{\frac{n}{1-n}} \label{eqn:dotnu}, 
\end{align}
where $C_1$ is an integration constant.  Taking the ratio $\nu/\dot{\nu}$ gives

\begin{align}
    \nu/\dot{\nu} = & -\frac{n-1}{k}(C_1 + kt), \label{eqn:nurat} \\
    \frac{C_1}{k} = & -\left(\frac{\nu}{\dot{\nu}(n-1)} + t \right). \label{eqn:c1k}
\end{align}
Solving Equation \ref{eqn:c1k} at $t = 939$ years with the Crab spin frequency $\nu$ = 30.2 Hz and spin frequency derivative $\dot{\nu}$ = -3.86 $\times$ 10$^{-10}$ Hz s$^{-1}$ \citep{Lyne1993} gives $C_1/k = 2.25 \times 10^{10}$ s. Substituting this for into Equation \ref{eqn:nu} at $t = 939$ years allows us to solve for $C_1$,

\begin{equation}
    C_1 = \frac{\nu^{1-n}}{n-1}\left( 1 + \frac{t}{C_1/k}\right)^{-1} = 1.7 \times 10^{-3} ,
    \label{eqn:c1}
\end{equation}
for n = 2.5.  Then, solving for $\nu$ at $t = 0$ gives the initial spin frequency 

\begin{equation}
    \nu_0 = \left( (n-1)C_1\right)^{\frac{1}{1-n}} = 53 \text{ Hz},
\end{equation}
corresponding to an initial spin period of 19 ms. Repeating the above calculation with $n = 3$ gives a spin frequency of 61 Hz, or spin period of 16 ms. 

The initial pulsar conditions derived here would have a luminosity too low to power the observed supernova light curve, implying that the assumption of constant $k$ and $n$ is incorrect if the pulsar-driven scenario is correct.  This shows the importance of accounting for the historical supernova data when estimating the initial conditions of the Crab pulsar.



\bsp	
\label{lastpage}
\end{document}